\documentclass[prd,aps,nofootinbib,superscriptaddress,twocolumn]{revtex4}

\usepackage{graphicx}
\usepackage{epsfig}
\begin{document}
\preprint{ANL-HEP-PR-07-53}

\title{Transverse momentum dependence of the angular distribution\\ 
       of the Drell-Yan process}

\author{Edmond L. Berger}
\email{berger@anl.gov}
\affiliation{High Energy Physics Division, 
      Argonne National Laboratory, 
      Argonne, IL 60439, U.S.A.}
\author{Jian-Wei Qiu} 
\email{jwq@iastate.edu}
\affiliation{High Energy Physics Division, 
      Argonne National Laboratory, 
      Argonne, IL 60439, U.S.A.}
\affiliation{Department of Physics and Astronomy, 
      Iowa State University,
      Ames, IA 50011, U.S.A.}
\author{Ricardo A. Rodriguez-Pedraza}
\email{rirodri@iastate.edu}
\affiliation{Department of Physics and Astronomy, 
      Iowa State University,
      Ames, IA 50011, U.S.A.}

\begin{abstract}
We calculate the transverse momentum $Q_\perp$ dependence of the helicity 
structure functions for the hadroproduction of a massive pair of leptons 
with pair invariant mass $Q$. These structure functions determine the 
angular 
distribution of the leptons in the pair rest frame.  Unphysical behavior in 
the region $Q_{\perp} \rightarrow 0$ is seen in the results of calculations 
done at fixed-order in QCD perturbation theory.  We use current conservation 
to demonstrate that the unphysical inverse-power and $\ln(Q/Q_{\perp})$ 
logarithmic divergences in three of the four independent
helicity structure functions share the same 
origin as the divergent terms in fixed-order calculations of the 
angular-integrated cross section.  We show that the resummation of these 
divergences to all orders in the strong coupling strength $\alpha_s$ can 
be reduced to the solved problem of the resummation of the divergences in   
the angular-integrated cross section, resulting in well-behaved predictions in 
the small $Q_{\perp}$ region. Among other results, we show the resummed 
part of the helicity structure functions preserves the Lam-Tung relation 
between the longitudinal and double spin-flip structure functions as a 
function of $Q_{\perp}$ to all orders in $\alpha_s$. 
\end{abstract}
\date{\today}


\maketitle
\section{Introduction}

Production of a massive pair of leptons of 
opposite electric charge in hadronic 
interactions, $h_1 + h_2 \rightarrow \ell^+ \ell^- X$, 
has revealed new narrow hadronic 
states, notably the $J/\Psi$ and the $\Upsilon$, 
and it continues to provide an  
important complement to deep-inelastic lepton scattering 
and other hard-scattering 
processes for probing the short-distance dynamics of strong 
and electroweak interactions.  
The assumption that the broad continuum of 
$\ell^+ \ell^-$ pairs originates from 
quark-antiquark annihilation through a single virtual photon, 
as embodied in the 
Drell-Yan model~\cite{Drell:1970wh}, 
implies that the angular distribution in the 
$\ell^+ \ell^-$ rest frame 
should be that of a transversely polarized photon, 
$(1 + \cos^2 \theta)$, where the 
polar angle $\theta$ is the direction of the lepton 
relative to the direction of the 
incident quark and antiquark.  
Acceptance restrictions limit measurements of the full 
angular distribution, but qualitative verification of 
this expectation was one 
of the early tests that increased confidence 
in the model~\cite{earlyangulardata}.  

In practice, massive lepton pairs are produced with 
substantial transverse momentum 
$Q_{\perp}$, supplied from a theoretical perspective by higher-order processes 
in perturbative quantum chromodynamics (QCD).  
An interesting challenge has been to predict 
how the angular distribution should behave 
as a function of $Q_{\perp}$
\cite{bdw-dy,lt-dy,Kajantie:1978yp,Cleymans:1978je,lt-dy2,Collins:1978yt,nlo,Chiappetta:1986yg,bv-dy}.
Indeed, this challenge is part of the more general ambition 
to predict the fully differential 
cross section $d \sigma/d Q d Q_{\perp} d y d \Omega$, 
where $Q$ is the invariant mass of 
the lepton pair, $y$ is its rapidity, and 
$d \Omega = d \cos \theta d \phi$ represents 
the differential decay angular distribution 
in the pair rest frame with respect to a suitably 
chosen set of axes. 

In addition to the virtual photon, the $W$ boson and the $Z$ boson also 
have important decay modes into pairs of leptons.  The angular distribution
of these leptons, measured in the rest-frame of the parent states,
determines the alignment (polarization) of the vector boson
and, consequently, supplies more precise information on the production 
dynamics than is accessible from the spin-averaged rate alone.  An 
understanding of the changes expected in the angular distribution as 
a function of the transverse momentum $Q_{\perp}$ is a topic of 
considerable importance, both for refined tests of QCD and for 
electroweak precision measurements.  An example of a QCD process is 
the flavor dependence of $W$ production in polarized hadron-hadron 
scattering at the Brookhaven Relativistic Heavy Ion Collider 
(RHIC)~\cite{spin-whitepaper}.
Better understanding of the expected angular distributions will reduce 
the systematic uncertainties on the determination of the W boson 
mass \cite{bqy-wz,Ellis:1997sc} and, in turn, improve the bound on the 
mass of the Higgs boson within the standard model of particle physics.  

\begin{figure}[t!]
\begin{center} 
\psfig{file=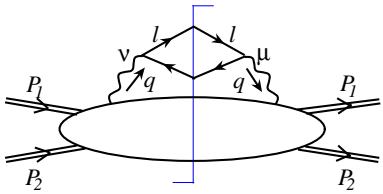,width=2.0in,angle=0}
\caption{Diagrammatic representation of hadronic dilepton production
via a virtual photon of four-momentum $q$. 
}
\label{fig1:dy}
\end{center} 
\end{figure}

In this paper we consider the scattering of two hadrons of momentum 
$P_1$ and $P_2$, respectively, 
producing a virtual photon of four-momentum $q$, 
$A(P_1)+B(P_2)\rightarrow \gamma^*(q)+X$, 
that decays into a pair of leptons of momentum $l$ and $\bar{l}$, as
sketched in Fig.~\ref{fig1:dy}.  The ideas and techniques developed here 
can be applied readily to the production of $W$ and $Z$ bosons, as well as 
to other yet-to-be-observed massive vector bosons that decay into 
a pair of leptons.  They are applicable also in semi-inclusive deep-inelastic 
scattering (SIDIS).  
 
The general formalism for the description of the angular 
distribution in terms of helicity structure functions is developed for the
Drell-Yan process in Ref.~\cite{lt-dy}.  The differential cross section may be
expressed as \cite{lt-dy}
\begin{eqnarray}
\frac{d\sigma}{d^4q d\Omega}
&=&
\frac{\alpha_{\rm em}^2}{2(2\pi)^4 S^2 Q^2}
\left[W_T (1+\cos^2\theta) \right.
\nonumber\\
&& 
+ W_L (1-\cos^2\theta) + W_{\Delta} (\sin(2\theta)\cos\phi) 
\nonumber\\
&& 
\left.
+ W_{\Delta\Delta} (\sin^2\theta\cos(2\phi)) \right]\, .
\label{x-sec-angular}
\end{eqnarray}
The four independent ``helicity'' structure functions $W_T$, $W_L$,  
$W_{\Delta}$, and $W_{\Delta \Delta}$ depend on $Q$, $Q_\perp$, rapidity 
$y$, and on the center-of-mass energy $\sqrt{S}$ of the 
production process.  They are defined in the virtual photon's
rest frame. and they correspond, respectively, to the transverse spin, 
longitudinal spin, single spin-flip, and double spin-flip contributions 
to the Drell-Yan cross section.  

The angular-integrated cross section is expressed 
in terms of $W_T$ and $W_L$ as 
\begin{eqnarray}
\frac{d\sigma}{d^4q} = \frac{\alpha_{\rm em}^2}{12 \pi^3 S^2 Q^2}
\left[2 W_T  + W_L \right]\ .
\label{x-sec-integ}
\end{eqnarray}
An interesting relationship $W_L = 2 W_{\Delta \Delta}$ 
between the longitudinal 
and double-flip structure functions is derived 
in Ref.~\cite{lt-dy} in the context of 
the parton model, and it has been shown to hold at least approximately 
at higher orders in perturbative QCD.  
Experimental tests of this relationship are reported in 
Ref.~\cite{Falciano:1986wk,Guanziroli:1987rp,Conway:1989fs,Heinrich:1991zm,Zhu:2006gx}.

Our principal focus in this paper is the prediction 
of the full $Q_{\perp}$ dependence of the four structure functions, including  
the region of small and intermediate $Q_{\perp}$ 
where the cross section takes on 
its largest values.  
Many papers dealing with various aspects of Drell-Yan angular 
distributions have preceded ours.  Explicit perturbative calculations
were done in the parton model \cite{bdw-dy,lt-dy}, in
perturbative QCD at order $\alpha_s$ 
\cite{Kajantie:1978yp,Collins:1978yt,Cleymans:1978je,lt-dy2}
and $\alpha_s^2$ \cite{nlo}, as well as in high twist formalisms
\cite{Berger:1979du,Qiu:1990xx}.
When calculated at fixed order in QCD perturbation theory, 
the structure functions show unphysical inverse-power 
$Q_\perp^{-n}$ ($n = 1$ or $2$) or logarithmic $\ln (Q/Q_{\perp})$ 
divergences, or both, as $Q_{\perp} \rightarrow 0$.  
For the angular-integrated cross section, $d\sigma/d^4q$, it is 
well established that similar unphysical divergences can be removed
after resummation of the $\ln^m(Q^2/Q_\perp^2)/Q_\perp^2$  
singular terms from initial-state gluon
emission to all orders in $\alpha_s$ 
\cite{DDT-qt,pp-b,cs-b,css-resum}. 

Examinations of the singular logarithmic terms in the 
helicity structure functions are reported in 
Refs.~\cite{Chiappetta:1986yg,bqy-wz,Ellis:1997sc,bv-dy}. 
Since only $W_T$ shows the $\ln^m(Q^2/Q_\perp^2)/Q_\perp^2$ 
divergence, previous resummation calculations were carried out only 
for $W_T$ in the same way as for the angular-integrated
cross section.  
As shown in Refs.~\cite{Chiappetta:1986yg,bqy-wz,Ellis:1997sc},
resummation removes the perturbative power divergence in 
$W_T$.  One consequence of resummation of just $W_T$ is a large 
change in the relative size of $W_T$ and the helicity structure 
functions for which no resummation is performed.  This result is not quite 
consistent with general expectations about the relative size of 
helicity structure functions in the Collins-Soper frame.  
For example, one expects $W_{\Delta\Delta}/W_T \rightarrow Q_\perp^2$ as
$Q_\perp\to 0$ \cite{cs-frame}.  

In Ref.~\cite{bv-dy}, Boer and Vogelsang carefully
investigate the logarithmic behavior of the order $\alpha_s$ 
perturbative contributions to the helicity structure functions.
At order $\alpha_s$, they find that, like $W_T$, 
both $W_L$ and $W_{\Delta\Delta}$ have a $\ln(Q^2/Q_\perp^2)$ 
logarithmic divergence, but not the $1/Q_\perp^2$ power divergence 
seen in $W_T$, and that $W_\Delta$ has no logarithmic divergence at 
this order in the Collins-Soper frame.  
They notice that the logarithmic contribution to
$W_L$ and $W_{\Delta\Delta}$ from quark-gluon (or gluon-quark)
subprocess is different from that for $W_T$ and does not fit 
the pattern expected for the perturbative expansion of 
the Collins-Soper-Sterman resummation
formalism to order $\alpha_s$ \cite{css-resum}.  
They also discuss the frame dependence of this logarithmic contribution.

The present paper expands on our earlier short manuscript on the same 
subject \cite{bqr-short}.  We start with 
the observations that the four helicity structure functions cannot 
be independent at $Q_\perp=0$ and that the general 
tensor decomposition in the virtual photon rest frame in
Eq.~(\ref{W-q0sfs}) is ill-defined at $Q_\perp=0$.  Then, guided 
by electromagnetic current conservation, we construct a new 
asymptotic form for the hadronic tensor 
with the right degrees of freedom as $Q_\perp\to 0$.  
We find that the leading logarithmic behavior of the 
different helicity structure functions, $W_T$, $W_L$, and 
$W_{\Delta\Delta}$, has a unique origin. 
We reduce the problem of transverse momentum 
resummation for $W_T$, $W_L$, and $W_{\Delta \Delta}$ to 
the known solution of transverse momentum resummation 
for the angular-integrated cross section \cite{css-resum},
and we prove that the logarithmic divergences 
in $W_T$, $W_L$, and $W_{\Delta \Delta}$ may be 
resummed to all orders in the strong coupling strength $\alpha_s$, 
yielding well behaved predictions that satisfy the expected 
kinematic constraints at small $Q_{\perp}$.  
We emphasize three main results of our research: 
\begin{itemize}
\item
Current conservation uniquely ties the perturbative divergences as 
$Q_\perp/Q\to 0$ of the otherwise independent helicity structure functions 
$W_T, W_L,$ and $W_{\Delta\Delta}$ to the divergence 
of the angular-integrated cross section.  
\item
The perturbative divergence in the angular-integrated cross section 
is sufficient 
to remove all leading divergences of the 
four individual helicity structure functions.  
\item
Transverse momentum resummation of the 
angular-integrated cross section determines the resummation 
of the large logarithmic 
terms of the helicity structure 
functions $W_T$, $W_L$, and $W_{\Delta\Delta}$, and 
the approximate Lam-Tung relation is an all-orders consequence 
of current conservation for the leading 
perturbatively divergent terms. 
\end{itemize}

The rest of our paper is organized as follows.
In Sec.~II, we define the helicity structure functions, and 
we derive the QCD perturbative contributions at order $\alpha_s$.  
We work in this paper entirely in the context of 
collinear QCD factorization \cite{css-fac}, meaning 
that $Q_{\perp} > \Lambda_{QCD}$, although $Q_\perp/Q$ may be small.  We 
examine in detail the leading behavior of the perturbative contributions 
to the helicity structure functions in the limit of 
small $Q_\perp/Q$.  In Secs.~III and IV, we investigate the generic singular 
structure of the perturbative contribution to the Drell-Yan hadronic tensor, 
and we derive an asymptotic current-conserving tensor that explicitly 
includes all the leading divergences of the perturbatively calculated helicity 
structure functions in the limit $Q_\perp/Q\to 0$.  We also explore the 
connection between cross 
sections for incident parton states of fixed helicity and the subleading 
perturbative contribution to the spin-averaged helicity structure functions.
We discuss all-orders transverse momentum resummation for helicity structure
functions in Sec.~V obtaining well-behaved distributions as a function of 
$Q_\perp$.  We show that the resummed part of the helicity structure
functions satisfies the Lam-Tung relation, $W_L=2W_{\Delta\Delta}$, between 
the longitudinal and the double-spin-flip structure function 
to all orders in $\alpha_s$.  Finally, in Sec.~VI, we offer a summary and our 
conclusions, and we outline plans for future work on W and Z hadroproduction 
and in semi-inclusive deep-inelastic scattering.  Three appendices are included 
in which we present detailed technical derivations of points discussed in the 
main body of the text.  

\section{Helicity structure functions and perturbative contributions}

In this section, we define the helicity structure functions
of Eq.~(\ref{x-sec-angular}).  We present the next-to-leading order
perturbative contributions to these functions and examine the 
structure of the singular behavior of each helicity structure 
function $W_i$ as $Q_\perp/Q \rightarrow 0$.

\subsection{Definition and normalization}

Helicity structure functions are defined in the virtual photon's
rest frame.  Let $\epsilon_\lambda^\mu(q)$ be the virtual photon's 
polarization vector with three polarization states, 
$\lambda=\pm 1, 0$. The helicity structure functions are 
\begin{eqnarray}
W_T 
&=& 
W_{\mu\nu}\, \epsilon_1^{\mu*}\epsilon_1^\nu\, ,
\nonumber\\
W_L 
&=& 
W_{\mu\nu}\, \epsilon_0^{\mu*}\epsilon_0^\nu\, ,
\nonumber\\
W_{\Delta} 
&=& 
W_{\mu\nu} 
\left( \epsilon_1^{\mu*}\epsilon_0^\nu
     + \epsilon_0^{\mu*}\epsilon_1^\nu \right)/\sqrt{2}\, ,
\nonumber\\
W_{\Delta\Delta} 
&=& 
W_{\mu\nu}\, \epsilon_1^{\mu*}\epsilon_{-1}^\nu\, ,
\label{helicity-f-def}
\end{eqnarray}
for the transverse spin, longitudinal spin, single spin-flip, and 
double spin-flip contributions to the Drell-Yan cross section, 
respectively.  In the virtual photon rest frame (the 
center-of-mass frame of the dilepton pair), the polarization 
vectors can be expressed in terms of orthogonal unit vectors 
in that frame, $X^\mu$, $Y^\mu$, and $Z^\mu$, as
$\epsilon_\pm^\mu = (\mp X^\mu - i Y^\mu)/\sqrt{2}$, 
$\epsilon_0^\mu = Z^\nu$ \cite{lt-dy}.  These 
unit vectors are normalized as $X^2=Y^2=Z^2=-1$, and 
they are also orthogonal to the current vector $q^\mu$.  They 
conserve the current, 
$q_\mu X^\mu=q_\mu Y^\mu=q_\mu Z^\mu=0$.  
Naturally, we can choose the fourth unit vector for the
$\vec{q}=0$ Lorentz frame to be $T^\mu=q^\mu/Q$ with $T^2=1$ and  
$Q=\sqrt{q^2}$.  The full Drell-Yan hadronic tensor can be
written in terms of the helicity structure functions and unit
vectors in the virtual photon rest frame as \cite{lt-dy}
\begin{eqnarray}
W^{\mu\nu} 
&=& 
- \left(g^{\mu\nu}-T^\mu T^\nu\right) 
  \left(W_T + W_{\Delta\Delta}\right)
\nonumber \\
&&
- 2 X^\mu X^\nu W_{\Delta\Delta} 
+ Z^\mu Z^\nu \left( W_L - W_T - W_{\Delta\Delta} \right)
\nonumber \\
&&
- \left( X^\mu Z^\nu + X^\nu Z^\mu \right) W_{\Delta}\, . 
\label{W-q0sfs}
\end{eqnarray}
Different choices of the axes lead to different $\vec{q}=0$
frames~\cite{lt-dy}.  
We choose to work in the Collins-Soper frame \cite{cs-frame}, 
whose unit vectors are defined as, 
\begin{eqnarray}
Z^\mu &=& 
\frac{2}{\sqrt{Q^2+Q_\perp^2}}
\left[ q_{P_2}\, \widetilde{P}_1^\mu
      -q_{P_1}\, \widetilde{P}_2^\mu  \right]\, ,
\nonumber \\
X^\mu &=& 
- \left(\frac{Q}{Q_\perp}\right) 
\frac{2}{\sqrt{Q^2+Q_\perp^2}}
\left[ q_{P_2}\, \widetilde{P}_1^\mu
      +q_{P_1}\, \widetilde{P}_2^\mu  \right]\, ,
\nonumber\\
Y^\mu &=& \epsilon^{\mu\nu\alpha\beta}\,T_\nu Z_\alpha X_\beta , 
\label{cs-def}
\end{eqnarray}
where the dimensionless current conserving hadron momenta are  
$\widetilde{P}_i^\mu=[P_i^\mu -(P_i\cdot q)/q^2\, q^\mu]/\sqrt{S}$ 
with $i=1,2$, and $q_{P_i} \equiv P_i\cdot q/\sqrt{S}$ with $i=1,2$.  
We present our derivation and predictions on helicity structure
functions in this Collins-Soper frame.  Transformation of our results 
to other commonly used frames is simply a rotation around the 
$Y$-axis \cite{lt-dy,bv-dy}.

\begin{figure}[t!]
\begin{center} 
\psfig{file=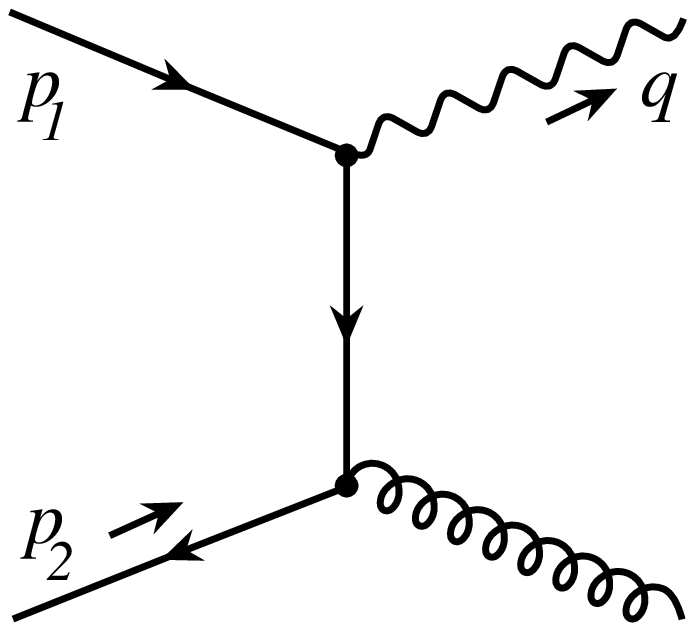,width=1.0in,angle=0}
\hskip 0.3in
\psfig{file=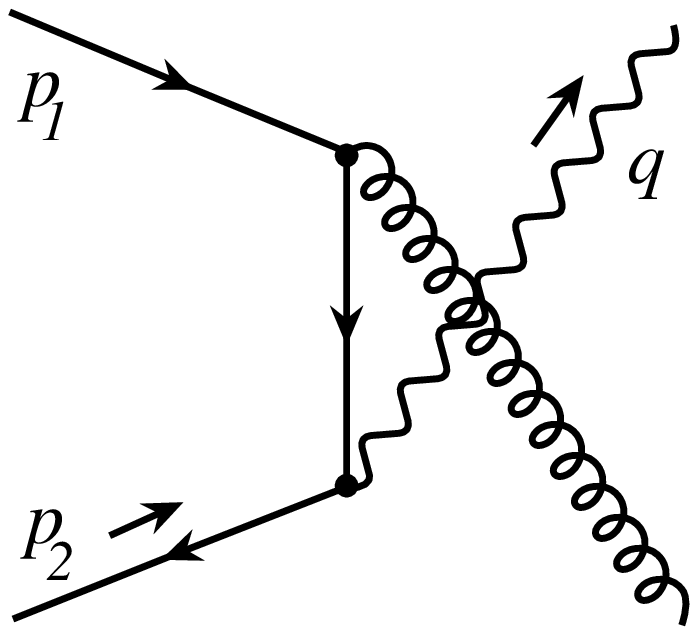,width=1.0in,angle=0}
\caption{Feynman diagrams for quark-antiquark annihilation to
a virtual photon plus a gluon. 
}
\label{fig2:qqb}
\end{center} 
\end{figure}

When the virtual photon mass $Q$ and its transverse momentum
$Q_\perp$ are much larger than $\Lambda_{\rm QCD}$, we expect 
QCD collinear factorization to be valid for the Drell-Yan cross 
section~\cite{css-fac}.  
Neglecting the transverse momentum of 
partons participating in the hard collisions, we write the incident
parton momenta as 
\begin{eqnarray}
p_1^\mu = \xi_1\, P_1^\mu \, ;
&\quad\quad&
p_2^\mu = \xi_2\, P_2^\mu \, .
\label{co-mom}
\end{eqnarray}
Neglecting all corrections suppressed by powers of  
$\Lambda_{\rm QCD}/Q$ or $\Lambda_{\rm QCD}/Q_\perp$, 
we can factor the hadronic tensor as
\begin{eqnarray}
W^{\mu\nu} 
&=&
\sum_{ab}
\int \frac{d\xi_1}{\xi_1}  \int \frac{d\xi_2}{\xi_2}\,
\phi_a(\xi_1)\, \phi_b(\xi_2)\,
\nonumber \\
&& \times
\omega^{\mu\nu}_{ab\to \gamma^* X}(\xi_1,\xi_2,q) \, ,
\label{Wmn-fac}
\end{eqnarray}
with incoming parton distributions $\phi_f(\xi)$ of flavor $f$ and
momentum fraction $\xi$.  The short-distance partonic tensor is 
\begin{eqnarray}
\omega^{\mu\nu}_{ab\to\gamma^* X} 
&=&
S\,\overline{\sum} \left| M^\mu_{ab\to\gamma^* X} \right|^*
                   \left| M^\nu_{ab\to\gamma^* X} \right|
\nonumber \\
&& \times
(2\pi)^4\, \delta^4(p_1+p_2-q-\sum_x p_x)
\nonumber \\
&& \times
\prod_x \frac{d^3p_x}{(2\pi)^3 2E_x}\, .
\label{wmn-p}
\end{eqnarray}
At the most basic level, a massive virtual photon arises from
quark-antiquark annihilation $q+\bar{q}\to \gamma^*$
in a collision of hadrons, and it is produced with $Q_\perp=0$.
The corresponding partonic tensor is 
\begin{eqnarray}
\omega^{\mu\nu}_{q\bar{q}\to\gamma^*} 
&=&
\frac{1}{3}\, e_q^2\, 
\left[\bar{n}^\mu n^\nu + n^\mu \bar{n}^\nu - g^{\mu\nu}
\right]
\nonumber\\
&& \times\,
\xi_1\xi_2\,\delta(\xi_1-x_1)\, \delta(\xi_2-x_2)
\nonumber\\
&& \times\,
(2\pi)^4\, S\, \delta^2(Q_\perp)\, ,
\label{wmn-lo}
\end{eqnarray}
with color factor 1/3 and fractional quark charge $e_q$.  The
unit vectors are $\bar{n}^\mu=\delta^{\mu +}$ 
and $n^\mu=\delta^{\mu -}$, and
\begin{equation}
x_1 = \frac{Q}{\sqrt{S}}{\rm e}^{y}\, , \quad \quad 
x_2 = \frac{Q}{\sqrt{S}}{\rm e}^{-y}\, .
\end{equation}
The lowest order helicity structure functions from
$q\bar{q}\to \gamma^*$ are 
\begin{eqnarray}
W_T^{(0)} 
&=& 
\sum_q \frac{1}{3}\, e_q^2\,
\phi_q(x_1)\, \phi_{\bar{q}}(x_2)\, 
(2\pi)^4\, S \, \delta^2(Q_\perp)\, ,
\nonumber\\
W_L^{(0)}
&=& 
W_\Delta^{(0)}
=
W_{\Delta\Delta}^{(0)}
= 0\, .
\label{wi-lo}
\end{eqnarray}
First-order gluon radiation supplies finite $Q_\perp$, 
through the quark-antiquark and quark-gluon
subprocesses, $q+\bar{q}\to\gamma^*+g$ and $q+g\to\gamma^*+q$,
as sketched in Figs.~\ref{fig2:qqb} and \ref{fig3:qg},
respectively.  Perturbatively, these
finite-order subprocesses yield {\it singular} differential cross
sections as a function of $Q_\perp$ in the limit $Q_\perp/Q \to
0$. For the angular-integrated cross section, $d\sigma/d^{4}q$,
it is well established that this unphysical divergence
can be removed after resummation of the singular terms from 
initial-state gluon emission to all orders 
in $\alpha_s$~\cite{css-resum}.  
The dependence of the helicity structure functions on $Q_\perp$ 
is our central focus in the rest of this manuscript.

\subsection{Order $\alpha_s$ contribution}

In this section we present explicit expressions for the contributions at order 
$\alpha_s$ to the four helicity structure functions from the two subprocesses 
$q \bar q \rightarrow \gamma^* g$ and $q g \rightarrow \gamma^* q$.  
Although some of the perturbative results are available
in the literature, we present for completeness 
in Appendix \ref{appendix-B} and \ref{appendix-C}, the details of the 
perturbative calculation in a consistent notation for the spin-averaged 
and ``polarized'' contributions to the parton-level helicity structure 
functions in the Collins-Soper frame.  

To better identify the analytic behavior as $Q_{\perp}/Q \rightarrow 0$ of 
the perturbative contributions, we express the results in terms of two new 
variables,
\begin{equation}
z_1 \equiv \frac{x_1}{\xi_1}\, ,
\quad\quad\quad
z_2 \equiv \frac{x_2}{\xi_2}\, .
\end{equation}
The parton-level Mandelstam variables 
defined in Eq.~(\ref{stu-hat}) in Appendix \ref{appendix-B} are 
expressed as
\begin{eqnarray}
\hat{s}
&=& 
\frac{Q^2}{z_1\, z_2}\, ,
\nonumber \\
\hat{t}
&=&
- \frac{Q_\perp^2}
       {1-z_2\sqrt{1+Q_\perp^2/Q^2}} \, ,
\nonumber \\
\hat{u}
&=&
- \frac{Q_\perp^2}
       {1-z_1\sqrt{1+Q_\perp^2/Q^2}} \, ,
\label{stu-z}
\end{eqnarray}
and  
\begin{eqnarray}
\frac{1}{\hat{t}\,\hat{u}} 
&=& 
\frac{1}{\hat{s}\, Q_\perp^2}\, ,
\nonumber \\
\frac{1}{\hat{s}(-\hat{t})} 
&=& 
\frac{1}{\hat{s}\, Q_\perp^2}\,
\left[1-z_2 \sqrt{1+Q_\perp^2/Q^2} \right] .
\nonumber
\end{eqnarray}

In the following subsections, we present our calculation for  
spin-averaged and polarized incident partons, with our specification 
of polarized states presented below.

\subsubsection{Spin averaged quark-antiquark annihilation}

As derived in Eq.~(\ref{qqbar-sf-cs}) in Appendix \ref{appendix-B}, 
the contribution to the parton-level 
helicity structure functions from the quark-antiquark 
annihilation subprocess, after averaging over the spins of 
the incident quark and antiquark, are 
\begin{eqnarray}
w^{q\bar{q}}_T 
&=& 
e_q^2\,\frac{8\pi^2\alpha_s}{3}\,
\left(\frac{Q^2}{Q_\perp^2}\right)
C_F\left[ z_1^2 + z_2^2 \right]\, 
(1+\frac{1}{2} \frac{Q_\perp^2}{Q^2})\, 
\nonumber \\
&& \hskip 0.4in
\times
\frac{S}{z_1 z_2}\,\delta(\hat{s}+\hat{t}+\hat{u}-Q^2)\, ,
\nonumber\\
w^{q\bar{q}}_L
&=& 
e_q^2\,\frac{8\pi^2\alpha_s}{3}\,
C_F\left[ z_1^2 + z_2^2 \right]\, 
\nonumber \\
&& \hskip 0.4in
\times
\frac{S}{z_1 z_2}\,\delta(\hat{s}+\hat{t}+\hat{u}-Q^2)\, ,
\nonumber\\
w^{q\bar{q}}_{\Delta\Delta}
&=& 
e_q^2\,\frac{8\pi^2\alpha_s}{3}\,
\frac{1}{2}\,
C_F\left[ z_1^2 + z_2^2 \right]\, 
\nonumber \\
&& \hskip 0.4in
\times
\frac{S}{z_1 z_2}\,\delta(\hat{s}+\hat{t}+\hat{u}-Q^2)
\nonumber\\
&=& \frac{1}{2}\, w^{q\bar{q}}_L\,\, ,
\nonumber\\
w^{q\bar{q}}_\Delta
&=& 
e_q^2\,\frac{8\pi^2\alpha_s}{3}\,
\left(\frac{Q}{Q_\perp}\right)
C_F\left[ z_1^2 - z_2^2 \right]\, 
\nonumber \\
&& \hskip 0.4in
\times
\frac{S}{z_1 z_2}\,\delta(\hat{s}+\hat{t}+\hat{u}-Q^2)\, .  
\label{qqbar-sf-cs-z}
\end{eqnarray}
The color factor is written as $4/9=(1/3)\times C_F$,  
with $(1/3)$ being the color factor for the lowest order contribution
in Eq.~(\ref{wmn-lo}), and $C_F=4/3$.  
With the exchange of $z_1$ and $z_2$ (or $\hat{t}$ and $\hat{u}$),
Eq.~(\ref{qqbar-sf-cs-z}) is also valid for the antiquark-quark
scattering subprocess, except for $w^{\bar{q}q}_\Delta$ 
which acquires an extra overall minus sign
that arises from the minus sign in the expression for $w_\Delta$ in
Eq.~(\ref{bqr2cs-p}).

The phase space $\delta$-function can also be expressed 
in terms of the new variables as 
\begin{eqnarray}
&&\frac{S}{z_1 z_2}\,\delta(\hat{s}+\hat{t}+\hat{u}-Q^2)
\nonumber \\
&& 
=\frac{1}{x_1 x_2}\ 
\delta\bigg(
(1-z_1\sqrt{1+Q_\perp^2/Q^2})
\nonumber\\
&& \hskip 0.6in
\times 
(1-z_2\sqrt{1+Q_\perp^2/Q^2}) 
-Q^2_\perp/\hat{s}
\bigg)\, .
\label{delta-z}
\end{eqnarray}

\begin{figure}[t!]
\begin{center} 
\psfig{file=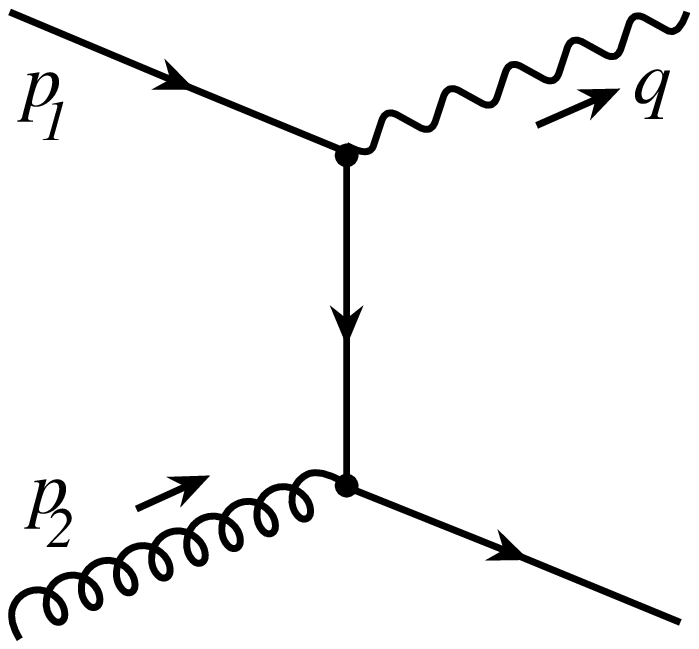,width=1.0in,angle=0}
\hskip 0.3in
\psfig{file=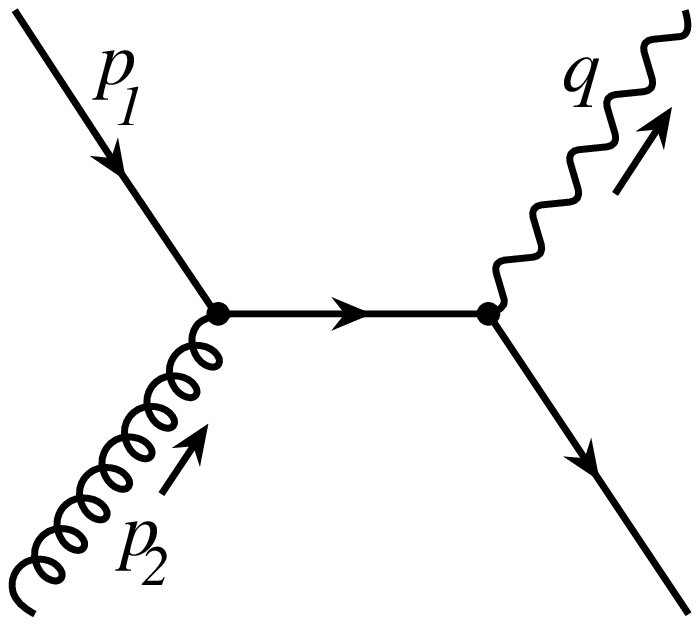,width=1.0in,angle=0}
\caption{Feynman diagrams for quark-gluon scattering to
produce a virtual photon plus a quark. 
}
\label{fig3:qg}
\end{center} 
\end{figure}

\subsubsection{Spin-averaged quark gluon scattering}

As derived in Eq.~(\ref{qg-sf-cs}) in Appendix \ref{appendix-B}, 
the contributions from the quark-gluon subprocess, after an 
average over the spins of the initial quark and gluon, are  
\begin{eqnarray}
w^{qg}_T 
&=& 
e_q^2\,\frac{8\pi^2\alpha_s}{3}
\left(\frac{Q^2}{Q_\perp^2}\right)
\left[1-z_2\sqrt{1+Q_\perp^2/Q^2}\right]
\nonumber\\
&& \hskip 0.3in
\times\, 
T_R\, \bigg[ \left[z_2^2 + (z_1 z_2-1)^2 \right]
\nonumber\\
&& \hskip 0.7in 
+\frac{1}{2}\,\frac{Q^2_\perp}{Q^2}
\left[z_2^2 - (z_1+z_2)^2 \right]\bigg]
\nonumber\\
&& \hskip 0.3in 
\times
\frac{S}{z_1 z_2}\,\delta(\hat{s}+\hat{t}+\hat{u}-Q^2)\, ,
\nonumber\\
w^{qg}_L
&=& 
e_q^2\,\frac{8\pi^2\alpha_s}{3}\,
\left[1-z_2\sqrt{1+Q_\perp^2/Q^2}\right]
\nonumber\\
&& \hskip 0.3in 
\times \,
T_R\,\left[z_2^2 + (z_1+z_2)^2 \right]
\nonumber \\
&& \hskip 0.3in 
\times
\frac{S}{z_1 z_2}\,\delta(\hat{s}+\hat{t}+\hat{u}-Q^2)\, ,
\nonumber\\
w^{qg}_{\Delta\Delta}
&=& 
e_q^2\,\frac{8\pi^2\alpha_s}{3}\,
\left[1-z_2\sqrt{1+Q_\perp^2/Q^2}\right]
\nonumber\\
&& \hskip 0.3in 
\times 
\frac{1}{2}\,T_R\,\left[z_2^2 + (z_1+z_2)^2 \right]
\nonumber \\
&& \hskip 0.3in 
\times
\frac{S}{z_1 z_2}\,\delta(\hat{s}+\hat{t}+\hat{u}-Q^2)
\nonumber\\
&=&\frac{1}{2}\,
w^{qg}_L ,
\nonumber\\
w^{qg}_\Delta
&=& 
e_q^2\,\frac{8\pi^2\alpha_s}{3}
\left(\frac{Q}{Q_\perp}\right)
\left[1-z_2\sqrt{1+Q_\perp^2/Q^2}\right]
\nonumber\\
&& \hskip 0.3in 
\times \,
T_R\,\left[z_1^2 - 2 z_2^2 \right]
\nonumber \\
&& \hskip 0.3in 
\times
\frac{S}{z_1 z_2}\,\delta(\hat{s}+\hat{t}+\hat{u}-Q^2)\, .
\label{qg-sf-cs-z}
\end{eqnarray}
The color factor is written as $1/6=(1/3)\times T_R$ with 
$T_R=1/2$.
With $z_1$ and $z_2$ switched, Eq.~(\ref{qg-sf-cs-z}) 
is also true for the gluon-quark scattering subprocesses, 
except for $w^{gq}_\Delta$ which acquires an extra overall minus sign
that arises from the minus sign in the expression for $w_\Delta$ in
Eq.~(\ref{bqr2cs-p}).

\subsubsection{Expressions for polarized incident partons}

The behavior at small $Q_{\perp}/Q$ of the parton-level 
helicity structure 
functions is sensitive to the helicity states of the incoming partons.  
We present here the perturbative contribution to the
helicity structure functions from the $q+\bar{q}\to\gamma^*+g$ 
and $q+g\to\gamma^*+q$ subprocesses with initial-state 
(anti)quark and gluon in a fixed helicity state.
For a quark of momentum $p$, the helicity projection
operator is
\begin{equation}
\widehat{P}_\pm(p) 
= \frac{1}{2}\,\gamma\cdot p\, 
  \pm \,
  \frac{1}{2}\,\gamma\cdot p \gamma_5 \, ,
\label{quark-proj}
\end{equation}
where the first term on the right-hand-side (RHS) corresponds to 
the projection for a spin-averaged quark state, while the second 
term corresponds to the projection for a polarized 
quark state, defined as the state with incoming quark polarization 
projected onto the {\it difference} of the quark's helicity states.
Similarly, the helicity projection operator
for a massless gluon of momentum $p$ moving in either the
light-cone ``$+$'' or ``$-$'' direction is 
\begin{equation}
P^{\alpha\beta}_\pm(p) 
= \frac{1}{2}\, d^{\alpha\beta}\, 
  \pm \,
  \frac{1}{2}\, i\, \epsilon^{\alpha\beta} ,
\label{gluon-proj}
\end{equation}
where the transverse tensor 
$d^{\alpha\beta}=-g^{\alpha\beta}+\bar{n}^\alpha n^\beta 
+ n^\alpha \bar{n}^\beta$,
$\epsilon^{\alpha\beta}=\epsilon^{\alpha\beta\rho\sigma} 
\bar{n}_\rho n_\sigma$.   
the first term on the RHS again corresponds to the 
projection for a spin-averaged and physically polarized gluon state, 
while the second term corresponds to the projection to 
a polarized gluon state, defined as the state with incoming 
gluon polarization projected onto the {\it difference} of 
the gluon's physically polarized states.

The contribution with a mixed unpolarized and a polarized parton state 
leads to an antisymmetric contribution to the hadronic tensor
$W_{\mu\nu}$, and it does not contribute to the Drell-Yan angular
distribution.  The sum or difference of our unpolarized and polarized 
contribution correspond to the contributions from initial-state
partons of the same or different fixed helicity state.  


Equation~(\ref{qqb-pol}) in Appendix \ref{appendix-C} shows 
that the polarized quark-antiquark contributions to the helicity
structure functions are the same as the unpolarized contributions,
\begin{eqnarray}
\Delta w^{q\bar{q}}_T 
&=& 
w^{q\bar{q}}_T\, , 
\nonumber\\
\Delta w^{q\bar{q}}_L
&=& 
w^{q\bar{q}}_L\, ,
\nonumber\\
\Delta w^{q\bar{q}}_{\Delta\Delta}
&=& 
w^{q\bar{q}}_{\Delta\Delta}
\nonumber \\
\Delta w^{q\bar{q}}_\Delta
&=& 
w^{q\bar{q}}_\Delta
\label{dqqbar-sf-cs-z}
\end{eqnarray}
with all unpolarized contributions given in 
Eq.~(\ref{qqbar-sf-cs-z}).


In treating quark-gluon scattering, we present results separately for 
the quark gluon and gluon quark initial states.  Equation~(\ref{dqg-sf-cs}) in 
Appendix \ref{appendix-C} provides the contribution 
from the quark-gluon scattering subprocess with polarized initial-states: 
\begin{eqnarray}
\Delta w^{qg}_T 
&=& 
e_q^2\,\frac{8\pi^2\alpha_s}{3}
\left(\frac{Q^2}{Q_\perp^2}\right)
\left[1-z_2\sqrt{1+Q_\perp^2/Q^2}\right]
\nonumber\\
&& \hskip 0.3in
\times\, 
T_R\, \bigg[ \left[ z_2^2 - (z_1 z_2-1)^2 \right]
\nonumber\\
&& \hskip 0.7in 
+\frac{1}{2}\,\frac{Q^2_\perp}{Q^2}
\left[z_2^2 + (z_1+z_2)^2 \right]\bigg]
\nonumber\\
&& \hskip 0.3in 
\times
\frac{S}{z_1 z_2}\,\delta(\hat{s}+\hat{t}+\hat{u}-Q^2)\, ,
\nonumber\\
\Delta w^{qg}_L
&=& 
e_q^2\,\frac{8\pi^2\alpha_s}{3}\,
\left[1-z_2\sqrt{1+Q_\perp^2/Q^2}\right]
\nonumber\\
&& \hskip 0.3in 
\times \,
T_R\,\left[z_2^2 -(z_1+z_2)^2 \right]
\nonumber \\
&& \hskip 0.3in 
\times
\frac{S}{z_1 z_2}\,\delta(\hat{s}+\hat{t}+\hat{u}-Q^2)\, ,
\nonumber\\
\Delta w^{qg}_{\Delta\Delta}
&=& 
e_q^2\,\frac{8\pi^2\alpha_s}{3}\,
\left[1-z_2\sqrt{1+Q_\perp^2/Q^2}\right]
\nonumber\\
&& \hskip 0.3in 
\times 
\frac{1}{2}\,T_R\,\left[z_2^2 -(z_1+z_2)^2 \right]
\nonumber \\
&& \hskip 0.3in 
\times
\frac{S}{z_1 z_2}\,\delta(\hat{s}+\hat{t}+\hat{u}-Q^2)
\nonumber\\
&=&\frac{1}{2}\,
\Delta w^{qg}_L
\nonumber\\
\Delta w^{qg}_\Delta
&=& 
e_q^2\,\frac{8\pi^2\alpha_s}{3}
\left(\frac{Q}{Q_\perp}\right)
\left[1-z_2\sqrt{1+Q_\perp^2/Q^2}\right]
\nonumber\\
&& \hskip 0.3in 
\times \,
T_R\,\left[ - z_1^2\right]
\nonumber \\
&& \hskip 0.3in 
\times
\frac{S}{z_1 z_2}\,\delta(\hat{s}+\hat{t}+\hat{u}-Q^2)\, .
\label{dqg-sf-cs-z}
\end{eqnarray}

As shown in Eq.~(\ref{dgq-sf-cs}) of Appendix \ref{appendix-C}, 
the contribution from the gluon-quark scattering subprocess with 
polarized initial-states is 
\begin{eqnarray}
\Delta w^{gq}_T 
&=& 
e_q^2\,\frac{8\pi^2\alpha_s}{3}
\left(\frac{Q^2}{Q_\perp^2}\right)
\left[1-z_1\sqrt{1+Q_\perp^2/Q^2}\right]
\nonumber\\
&& \hskip 0.3in
\times\, 
T_R\, \bigg[ \left[ z_1^2 - (z_1 z_2-1)^2 \right]
\nonumber\\
&& \hskip 0.7in 
+\frac{1}{2}\,\frac{Q^2_\perp}{Q^2}
\left[z_1^2 + (z_1+z_2)^2 \right]\bigg]
\nonumber\\
&& \hskip 0.3in 
\times
\frac{S}{z_1 z_2}\,\delta(\hat{s}+\hat{t}+\hat{u}-Q^2)\, ,
\nonumber\\
\Delta w^{gq}_L
&=& 
e_q^2\,\frac{8\pi^2\alpha_s}{3}\,
\left[1-z_1\sqrt{1+Q_\perp^2/Q^2}\right]
\nonumber\\
&& \hskip 0.3in 
\times \,
T_R\,\left[z_1^2 -(z_1+z_2)^2 \right]
\nonumber \\
&& \hskip 0.3in 
\times
\frac{S}{z_1 z_2}\,\delta(\hat{s}+\hat{t}+\hat{u}-Q^2)\, ,
\nonumber\\
\Delta w^{gq}_{\Delta\Delta}
&=& 
e_q^2\,\frac{8\pi^2\alpha_s}{3}\,
\left[1-z_1\sqrt{1+Q_\perp^2/Q^2}\right]
\nonumber\\
&& \hskip 0.3in 
\times 
\frac{1}{2}\,T_R\,\left[z_1^2 -(z_1+z_2)^2 \right]
\nonumber \\
&& \hskip 0.3in 
\times
\frac{S}{z_1 z_2}\,\delta(\hat{s}+\hat{t}+\hat{u}-Q^2)
\nonumber\\
&=&\frac{1}{2}\,
\Delta w^{gq}_L
\nonumber\\
\Delta w^{gq}_\Delta
&=& 
e_q^2\,\frac{8\pi^2\alpha_s}{3}
\left(\frac{Q}{Q_\perp}\right)
\left[1-z_1\sqrt{1+Q_\perp^2/Q^2}\right]
\nonumber\\
&& \hskip 0.3in 
\times \,
T_R\,\left[ z_2^2\right]
\nonumber \\
&& \hskip 0.3in 
\times
\frac{S}{z_1 z_2}\,\delta(\hat{s}+\hat{t}+\hat{u}-Q^2)\, .
\label{dgq-sf-cs-z}
\end{eqnarray}

We note that other than for $\Delta w^{gq}_\Delta$, 
the contributions from the gluon-quark subprocess are effectively
the same as those from the quark-gluon subprocess, with $z_1$ and $z_2$ 
switched.

\subsection{Limit of $Q_\perp/Q\to 0$}

In this subsection, we examine the analytic behavior of each parton-level 
helicity structure function as $Q_\perp/Q\to 0$.  
Keeping up to the leading power terms, we can simplify the parton-level 
Mandelstam variables and the phase space $\delta$-function as 
\begin{eqnarray}
\hat{s}
&\Rightarrow &
\frac{Q_\perp^2}
     {(1-z_1)(1-z_2)} \, ,
\nonumber \\
\hat{t}
&\Rightarrow &
- \frac{Q_\perp^2}
       {(1-z_2)} \, ,
\nonumber \\
\hat{u}
&\Rightarrow &
- \frac{Q_\perp^2}
       {(1-z_1)} \, .  
\label{stu-limit}
\end{eqnarray}
The expression for $\hat{s}$ is an immediate consequence
of the phase space $\delta$-function, which, in turn, can 
be expanded as \cite{mos-sidis}
\begin{eqnarray}
&&
\frac{S}{z_1 z_2}\, \delta(\hat{s}+\hat{t}+\hat{u}-Q^2)
\nonumber \\
&& \Rightarrow
\frac{1}{x_1 x_2}
\bigg[
\frac{\delta(1-z_2)}{(1-z_1)_+} + \frac{\delta(1-z_1)}{(1-z_2)_+}
\nonumber \\
&& \hskip 0.5in
 + \delta(1-z_1)\, \delta(1-z_2)\, \ln\frac{Q^2}{Q_\perp^2}
\bigg] .
\label{delta-qt0}
\end{eqnarray}
The standard definition of ``+'' distribution is 
\begin{equation}
\int_x^1 dz\, \frac{f(z)}{(1-z)_+}
= \int_x^1 dz\, \frac{f(z)-f(1)}{(1-z)} + f(1)\, \ln(1-x)
\label{plus-dis}
\end{equation}


Substituting Eqs.~(\ref{stu-limit}) and (\ref{delta-qt0}) into
Eqs.~(\ref{qqbar-sf-cs-z}) and (\ref{qg-sf-cs-z}), we obtain
the analytic behavior of the perturbatively calculated parton-level
helicity structure functions as $Q_\perp/Q\to 0$.  For the 
quark-antiquark annihilation process, these are  
\begin{eqnarray}
w^{q\bar{q}}_T 
&\Rightarrow & 
e_q^2\,\frac{8\pi^2\alpha_s}{3x_1x_2}
\left(\frac{Q^2}{Q_\perp^2}\right)
\bigg\{
P_{qq}(z_2)\delta(1-z_1)
\nonumber \\
&& \hskip 1.0in
+P_{qq}(z_1)\delta(1-z_2)
\nonumber \\
&& 
+2\, C_F\delta(1-z_1)\delta(1-z_2)
\left[\ln(\frac{Q^2}{Q_\perp^2})-\frac{3}{2}\right]
\bigg\}\, ,
\nonumber\\
w^{q\bar{q}}_L
&\Rightarrow & 
e_q^2\,\frac{8\pi^2\alpha_s}{3x_1x_2}
\bigg\{
P_{qq}(z_2)\delta(1-z_1)+P_{qq}(z_1)\delta(1-z_2)
\nonumber \\
&& 
+2\, C_F \delta(1-z_1)\delta(1-z_2)
\left[\ln(\frac{Q^2}{Q_\perp^2})-\frac{3}{2}\right]
\bigg\}\, ,
\nonumber\\
w^{q\bar{q}}_{\Delta\Delta}
&\Rightarrow & 
\frac{1}{2}\,
e_q^2\,\frac{8\pi^2\alpha_s}{3x_1x_2}
\bigg\{
P_{qq}(z_2)\delta(1-z_1)
\nonumber \\
&& \hskip 0.7in
+P_{qq}(z_1)\delta(1-z_2)
\nonumber \\
&& 
+2\, C_F \delta(1-z_1)\delta(1-z_2)
\left[\ln(\frac{Q^2}{Q_\perp^2})-\frac{3}{2}\right]
\bigg\}\, ,
\nonumber\\
w^{q\bar{q}}_\Delta
&\Rightarrow & 
e_q^2\,\frac{8\pi^2\alpha_s}{3x_1x_2}
\left(\frac{Q}{Q_\perp}\right)
\bigg\{
C_F [1+z_2]\delta(1-z_1)
\nonumber \\
&& 
\hskip 1.0in
- C_F [1+z_1]\delta(1-z_2)
\bigg\} .
\label{qqbar-sf-cs-qt0}
\end{eqnarray}

For the quark-gluon subprocess, the small $Q_{\perp}$ behavior is  
\begin{eqnarray}
w^{qg}_T 
&\Rightarrow & 
e_q^2\,\frac{8\pi^2\alpha_s}{3x_1x_2}
\left(\frac{Q^2}{Q_\perp^2}\right)
P_{qg}(z_2)\, \delta(1-z_1)\, ,
\nonumber\\
w^{qg}_L
&\Rightarrow & 
e_q^2\,\frac{8\pi^2\alpha_s}{3x_1x_2}\
P_{qg}(-z_2)\, \delta(1-z_1)\, ,
\nonumber\\
w^{qg}_{\Delta\Delta}
&\Rightarrow & 
\frac{1}{2}\,
e_q^2\,\frac{8\pi^2\alpha_s}{3x_1x_2}\
P_{qg}(-z_2)\, \delta(1-z_1)\, ,
\nonumber\\
w^{qg}_\Delta
&\Rightarrow & 
e_q^2\,\frac{8\pi^2\alpha_s}{3x_1x_2}
\left(\frac{Q}{Q_\perp}\right)
T_R \left[1-2z_2^2\right]\delta(1-z_1) .
\label{qg-sf-cs-qt0}
\end{eqnarray} 
The parton-to-parton splitting functions are
\begin{eqnarray}
P_{qq}(z) 
&=& C_F 
\left[ 
\frac{1+z^2}{(1-z)_+} + \frac{3}{2}\delta(1-z)
\right]\, ,
\label{pqq} \\
P_{qg}(z) 
&=& T_R 
\left[ 
z^2 + (1-z)^2
\right]\, .
\label{pqg}
\end{eqnarray}
With $z_1$ and $z_2$ switched, Eq.~(\ref{qg-sf-cs-qt0}) is also 
true for the gluon-quark subprocess, except for $w^{gq}_\Delta$ which
needs an extra overall minus sign.  Our results for the form of the 
helicity structure functions for unpolarized incoming partons 
as $Q_\perp/Q\to 0$ in Eqs.~(\ref{qqbar-sf-cs-qt0}) and 
(\ref{qg-sf-cs-qt0}) are consistent with those derived in
Ref.~\cite{bv-dy}.


Equation~(\ref{dqqbar-sf-cs-z}) allows us to conclude that, at this order, 
the analytic behavior of the quark-antiquark annihilation 
subprocess as $Q_\perp/Q\to 0$ is independent of whether incoming
(anti)quarks are spin-averaged or polarized. The contributions 
to the parton-level helicity structure functions are given in
Eq.~(\ref{qqbar-sf-cs-qt0}).

On the other hand, the polarized contributions from quark-gluon
scattering subprocess are different from those for ``spin-averaged'' initial
parton states.  From Eq.~(\ref{dqg-sf-cs-z}), we obtain
\begin{eqnarray}
\Delta w^{qg}_T 
&\Rightarrow & 
e_q^2\,\frac{8\pi^2\alpha_s}{3x_1x_2}
\left(\frac{Q^2}{Q_\perp^2}\right)
\Delta P_{qg}(z_2)\,\delta(1-z_1)
\nonumber\\
\Delta w^{qg}_L
&\Rightarrow & 
e_q^2\,\frac{8\pi^2\alpha_s}{3x_1x_2}\,
\Delta P_{qg}(-z_2)\,\delta(1-z_1)
\nonumber\\
\Delta w^{qg}_{\Delta\Delta}
&\Rightarrow & 
\frac{1}{2}\,
e_q^2\,\frac{8\pi^2\alpha_s}{3x_1x_2}\,
\Delta P_{qg}(-z_2)\,\delta(1-z_1)
\nonumber\\
\Delta w^{qg}_\Delta
&\Rightarrow & 
e_q^2\,\frac{8\pi^2\alpha_s}{3x_1x_2}
\left(\frac{Q}{Q_\perp}\right)
\left[-T_R\, \delta(1-z_1)\right] ,   
\label{dqg-sf-cs-qt0}
\end{eqnarray}
where $\Delta P_{qg}(z)$ is the leading polarized 
gluon-to-quark splitting function
\begin{equation}
\Delta P_{qg}(z) = T_R 
\left[ 
z^2 - (1-z)^2
\right]\, .
\label{dpqg}
\end{equation}
Similarly, based on Eq.~(\ref{dgq-sf-cs-z}), 
the small $Q_{\perp}$ behavior of the polarized gluon-quark 
contribution is  
\begin{eqnarray}
\Delta w^{gq}_T 
&\Rightarrow & 
e_q^2\,\frac{8\pi^2\alpha_s}{3x_1x_2}
\left(\frac{Q^2}{Q_\perp^2}\right)
\Delta P_{qg}(z_1)\,\delta(1-z_2)
\nonumber\\
\Delta w^{gq}_L
&\Rightarrow & 
e_q^2\,\frac{8\pi^2\alpha_s}{3x_1x_2}\,
\Delta P_{qg}(-z_1)\,\delta(1-z_2)
\nonumber\\
\Delta w^{gq}_{\Delta\Delta}
&\Rightarrow & 
\frac{1}{2}\,
e_q^2\,\frac{8\pi^2\alpha_s}{3x_1x_2}\,
\Delta P_{qg}(-z_1)\,\delta(1-z_2)
\nonumber\\
\Delta w^{gq}_\Delta
&\Rightarrow & 
e_q^2\,\frac{8\pi^2\alpha_s}{3x_1x_2}
\left(\frac{Q}{Q_\perp}\right)
\left[T_R\, \delta(1-z_2)\right]\, .
\label{dgq-sf-cs-qt0}
\end{eqnarray}

Clearly, the perturbatively calculated 
helicity structure functions at 
order of $\alpha_s$ and beyond are singular as $Q_\perp/Q\to 0$:
$W_T$ and $W_\Delta$ have the power divergences, 
$Q^2/Q_\perp^2$ and $Q/Q_\perp$, respectively, 
as well as $\ln(Q/Q_\perp)$ divergences,
whereas $W_L$ and $W_{\Delta\Delta}$ show 
$\ln(Q/Q_\perp)$ divergences 
\cite{Chiappetta:1986yg,bqy-wz,Ellis:1997sc,bv-dy}.

\section{Asymptotic current conserving tensor}

In this section, we investigate the possible connection between 
the logarithmic divergences of different helicity structure 
functions, and we show that they have a common origin.  We observe 
that the four helicity structure functions cannot be independent 
as $Q_\perp=0$ where the general tensor decomposition in the virtual 
photon rest frame in Eq.~(\ref{W-q0sfs}) is ill-defined.  
We construct a new asymptotic hadronic tensor that has the right 
number of independent scalar functions as $Q_\perp\to 0$ by requiring 
that the singular contribution to the hadronic tensor should satisfy 
electromagnetic current conservation to all orders in $\alpha_s$.
We show explicitly  
that the $Q_\perp/Q \rightarrow 0$ singular contributions 
in $W_T$, $W_L$, and $W_{\Delta\Delta}$ are related uniquely 
to the singular contribution of the angular-integrated cross section.

The general arguments in Ref.~\cite{cs-frame} show that there should be
only two independent power-divergent scalar functions as $Q_\perp/Q \rightarrow 0$ 
in the Collins-Soper frame.  To display the explicit dependence of the hadronic tensor 
on $Q_\perp/Q$, we rewrite the unit
vectors of the Collins-Soper frame in Eq.~(\ref{cs-def}) as
\begin{eqnarray}
T^\mu 
&=& 
\frac{1}{\sqrt{2}}\sqrt{1+\frac{Q_\perp^2}{Q^2}}
\left[ e^y\, \bar{n}^\mu + e^{-y}\, n^\mu \right]
+\left(\frac{Q_\perp}{Q}\right) n_\perp^\mu \, ,
\nonumber\\
Z^\mu 
&=& 
\frac{1}{\sqrt{2}}
\left[ e^y\, \bar{n}^\mu - e^{-y}\, n^\mu \right]\, ,
\label{cs-lab}\\
X^\mu 
&=& 
\frac{1}{\sqrt{2}}
\left(\frac{Q_\perp}{Q}\right) 
\left[ e^y\, \bar{n}^\mu + e^{-y}\, n^\mu \right]
+\sqrt{1+\frac{Q_\perp^2}{Q^2}}\, n_\perp^\mu \, ,
\nonumber 
\end{eqnarray}
with $Y^\mu$ uniquely fixed.  
By expanding the full Drell-Yan hadronic tensor in Eq.~(\ref{W-q0sfs})
and using Eq.~(\ref{cs-lab}) in the limit $Q_\perp/Q\to 0$, 
we obtain the following form for the singular terms 
of the tensor \cite{rr-thesis,bqr-short}
\begin{eqnarray}
W^{\mu\nu}_{\rm Sing}
&=& 
\left(-g^{\mu\nu}+\bar{n}^\mu n^\nu + n^\mu \bar{n}^\nu
\right) W_2^{\rm Asym}
\nonumber \\
&& + 
\frac{1}{\sqrt{2}} 
\left[ \frac{Q_\perp}{Q} 
       \left(n_\perp^\mu \bar{n}^\nu +
             \bar{n}^\mu n_\perp^\nu\right)\, {\rm e}^y 
\right] 
\nonumber \\
&& \hskip 0.2in \times
\left(W_2^{\rm Asym} - \frac{Q}{Q_\perp} W_1^{\rm Asym} \right)
\nonumber\\
&& + 
\frac{1}{\sqrt{2}} 
\left[ \frac{Q_\perp}{Q} 
       \left(n_\perp^\mu n^\nu +
             n^\mu n_\perp^\nu\right)\, {\rm e}^{-y} 
\right] 
\nonumber \\
&& \hskip 0.2in \times
\left(W_2^{\rm Asym} + \frac{Q}{Q_\perp} W_1^{\rm Asym} \right) .  
\label{w-singular}
\end{eqnarray}
At this point, there are two unspecified divergent scalar functions:
$W_2^{\rm Asym} \propto Q^2/Q_\perp^2$ and 
$W_1^{\rm Asym} \propto Q/Q_\perp$ as 
$Q_\perp/Q\to 0$. 
In Eq.~(\ref{w-singular}), the unit vectors $\bar{n},n,n_\perp$ 
specify the center-of-mass frame of the hadron collision, 
defined in Appendix \ref{appendix-A}.

The singular tensor as $Q_\perp/Q\to 0$ in Eq.~(\ref{w-singular}) is 
not current conserving since $q_\mu\, W^{\mu\nu}_{\rm Sing} \neq 0$. 
In order to 
resum the singular terms of the hadronic tensor to all orders
in $\alpha_s$, we require a tensor that incorporates 
all the singular terms {\em and} also conserves the current
perturbatively at any order of $\alpha_s$.  We use the term  
{\em asymptotic tensor} for this current-conserving tensor.  
We define it to be 
\begin{eqnarray}
W^{\mu\nu}_{\rm Asym}
&=& 
\left(-g^{\mu\nu}+\bar{n}^\mu n^\nu + n^\mu \bar{n}^\nu
\right) W_2^{\rm Asym}
\nonumber \\
&& + 
\frac{Q_\perp}{Q^-} 
\left(n_\perp^\mu \bar{n}^\nu + \bar{n}^\mu n_\perp^\nu
     +\frac{Q_\perp}{Q^-} \bar{n}^\mu \bar{n}^\nu \right) 
\nonumber \\
&& \hskip 0.2in \times
\frac{1}{2}
\left[W_2^{\rm Asym} - \frac{Q}{Q_\perp} W_1^{\rm Asym} \right]
\nonumber\\
&& + 
\frac{Q_\perp}{Q^+} 
\left(n_\perp^\mu n^\nu + n^\mu n_\perp^\nu
     +\frac{Q_\perp}{Q^+} n^\mu n^\nu \right) 
\nonumber \\
&& \hskip 0.2in \times
\frac{1}{2}
\left[W_2^{\rm Asym} + \frac{Q}{Q_\perp} W_1^{\rm Asym} \right]\, ,
\label{w-asym}
\end{eqnarray}
where the components of the virtual photon momentum $Q^+ = q \cdot n$ and 
$Q^- = q \cdot \bar{n}$ 
are defined in Appendix \ref{appendix-A}.
The asymptotic tensor in Eq.~(\ref{w-asym}) 
is equal to the singular tensor in Eq.~(\ref{w-singular}) 
plus a minimal non-singular term such that  
$q_\mu\, W^{\mu\nu}_{\rm Asym} = 0$.

The angular-integrated cross section is obtained from the trace,
$d\sigma/d^4q \propto -g_{\mu\nu}\,W^{\mu\nu}$.   The trace of the
asymptotic tensor in Eq.~(\ref{w-asym}) should therefore be fixed by 
the asymptotic term $W^{\rm Asym}$ of the angular-integrated
Drell-Yan transverse momentum distribution~\cite{css-resum}. This 
statement allows us to fix uniquely the asymptotically 
divergent function $W_2^{\rm Asym}$ in Eq.~(\ref{w-asym}). We obtain 
\begin{equation}
W_2^{\rm Asym} = W^{\rm Asym}/2 .  
\end{equation}
The angular-integrated cross section fixes the value of $W_2^{\rm Asym}$, but 
it cannot fix the second scalar function $W_1^{\rm Asym}$ in 
Eq.~(\ref{w-asym}). This second function represents the singular perturbative 
behavior of the structure function $W_\Delta$. We defer discussion of 
$W_\Delta$ until 
Sec.~V and concentrate on transverse momentum resummation for the other three 
helicity structure functions, $W_T$, $W_L$, and $W_{\Delta\Delta}$.

We reexpress the asymptotic tensor in terms of the previously 
defined unit vectors in the Collins-Soper frame as 
\begin{eqnarray}
W^{\mu\nu}_{\rm Asym}
&=& 
\left[
\left(-g^{\mu\nu}+T^\mu T^\nu\right)
-\frac{Q_\perp^2/Q^2}{1+Q_\perp^2/Q^2}\, X^\mu X^\nu
\right. \nonumber \\
&& \hskip 0.2in \left.
-\frac{1}{1+Q_\perp^2/Q^2}\, Z^\mu Z^\nu
\right]
\frac{W^{\rm Asym}}{2}
\nonumber \\
& - & 
\frac{1}{1+Q_\perp^2/Q^2}\,
\left[ X^\mu Z^\nu + Z^\mu X^\nu \right]
W_1^{\rm Asym}\, .
\label{w-asym-cs}
\end{eqnarray}
Upon comparison with Eq.~(\ref{W-q0sfs}), we immediately derive the 
corresponding asymptotic 
helicity structure functions,
\begin{eqnarray}
W_T^{\rm Asym}
&=& 
\left(1-\frac{1}{2} \frac{Q_\perp^2/Q^2}{1+Q_\perp^2/Q^2} \right)
\frac{W^{\rm Asym}}{2}
\approx
\frac{W^{\rm Asym}}{2}\, ,
\nonumber \\
W_L^{\rm Asym}
&=& 
\frac{Q_\perp^2/Q^2}{1+Q_\perp^2/Q^2}\,
\frac{W^{\rm Asym}}{2}
\approx
\frac{Q_\perp^2}{Q^2}\,
\frac{W^{\rm Asym}}{2}\, ,
\label{st-asym-cs} \\
W_{\Delta\Delta}^{\rm Asym}
&=& 
\frac{1}{2}\,\frac{Q_\perp^2/Q^2}{1+Q_\perp^2/Q^2}\,
\frac{W^{\rm Asym}}{2}
\approx
\frac{1}{2}\,\frac{Q_\perp^2}{Q^2}\,
\frac{W^{\rm Asym}}{2}\, .
\nonumber
\end{eqnarray}

Equation (\ref{st-asym-cs}) 
shows that current conservation relates the $Q_{\perp}/Q \rightarrow 
0$ divergent terms of the transverse, longitudinal, and 
double-spin flip structure functions intimately to the divergent 
part of the angular-integrated transverse momentum distribution.
The next key question, addressed affirmatively in the next section, 
is whether the asymptotic helicity structure functions
in Eq.~(\ref{st-asym-cs}), as derived here, are sufficient to remove 
all the leading divergences in the perturbatively calculated structure 
functions order by order in $\alpha_s$.

\section{Perturbative finite tensor}

We show in this section that the three asymptotic  
helicity structure functions presented in the last section include 
all the  $Q_{\perp}/Q \rightarrow 0$ leading divergent terms 
of the corresponding 
perturbatively calculated helicity structure functions, and 
therefore, that we can define a perturbatively finite tensor 
from the difference 
\begin{equation}
W^{\mu\nu}_{\rm Finite} \equiv
W^{\mu\nu}_{\rm Pert} - W^{\mu\nu}_{\rm Asym}\, ,
\label{w-finite}
\end{equation}
at any order of $\alpha_s$.  This finite tensor
conserves the current since the asymptotic tensor conserves
the current.

The  $Q_{\perp}/Q \rightarrow 0$ divergent part of the 
angular-integrated cross section is obtained from the trace of 
the hadronic tensor $g_{\mu\nu}\, W^{\mu\nu}$. Applying this 
statement at the parton level, we use the results 
of Sec.~III to derive the $Q_{\perp}/Q \rightarrow 0$ asymptotic  
terms for the angular-integrated and spin-averaged 
$q\bar{q} \rightarrow \gamma^* g$ and $qg \rightarrow \gamma^* q$ 
subprocesses.  These are 
\begin{eqnarray}
\frac{w_{q\bar{q}}^{\rm Asym}}{2}
&\approx & 
e_q^2\,\frac{8\pi^2\alpha_s}{3x_1x_2}\,\frac{Q^2}{Q_\perp^2}
\bigg\{
P_{qq}(z_2)\delta(1-z_1)
\nonumber \\
&& \hskip 0.9in
+P_{qq}(z_1)\delta(1-z_2)
\nonumber \\
&& 
+2\, C_F\delta(1-z_1)\delta(1-z_2)
\left[\ln(\frac{Q^2}{Q_\perp^2})-\frac{3}{2}\right]
\bigg\}\, ;
\nonumber\\
\frac{w_{qg}^{\rm Asym}}{2} 
&\approx & 
e_q^2\,\frac{8\pi^2\alpha_s}{3x_1x_2}\,\frac{Q^2}{Q_\perp^2}\,
P_{qg}(z_2)\,\delta(1-z_1)\, ;
\nonumber\\
\frac{w_{gq}^{\rm Asym}}{2} 
&\approx & 
e_q^2\,\frac{8\pi^2\alpha_s}{3x_1x_2}\,\frac{Q^2}{Q_\perp^2}\,
P_{qg}(z_1)\,\delta(1-z_2)\, .
\label{w-asym-unpol}
\end{eqnarray}

Using Eq.~(\ref{st-asym-cs}) at the parton level, we find that 
as $Q_\perp/Q\to 0$, 
the parton-level asymptotic terms in Eq.~(\ref{w-asym-unpol}) 
remove all divergent contributions of the corresponding 
perturbatively calculated helicity structure functions. For 
the quark-antiquark annihilation subprocess,  
\begin{eqnarray}
w^{q\bar{q}}_T 
&-& 
\left(1-\frac{1}{2} \frac{Q_\perp^2/Q^2}{1+Q_\perp^2/Q^2} \right)
\frac{w_{q\bar{q}}^{\rm Asym}}{2}\
\Rightarrow \,
{\cal O}(Q_\perp^0)
\nonumber \\
w^{q\bar{q}}_L 
&-&
\frac{Q_\perp^2/Q^2}{1+Q_\perp^2/Q^2}\,
\frac{w_{q\bar{q}}^{\rm Asym}}{2}\
\Rightarrow \,
{\cal O}(Q_\perp^2)
\nonumber \\
w^{q\bar{q}}_{\Delta\Delta} 
&-& \frac{1}{2}\,
\frac{Q_\perp^2/Q^2}{1+Q_\perp^2/Q^2}\,
\frac{w_{q\bar{q}}^{\rm Asym}}{2}\
\Rightarrow \,
{\cal O}(Q_\perp^2) . 
\label{sf-finite-qq-qt0}
\end{eqnarray}
For the quark-gluon subprocess,  
\begin{eqnarray}
w^{qg}_T 
&-& 
\left(1-\frac{1}{2} \frac{Q_\perp^2/Q^2}{1+Q_\perp^2/Q^2} \right)
\frac{w_{qg}^{\rm Asym}}{2}
\Rightarrow 
{\cal O}(Q_\perp^0)
\nonumber \\
w^{qg}_L 
&-& 
\frac{Q_\perp^2/Q^2}{1+Q_\perp^2/Q^2}\,
  \frac{w_{qg}^{\rm Asym}}{2}\
\Rightarrow \,
e_q^2\,\frac{8\pi^2\alpha_s}{3x_1x_2}\, \delta(1-z_1)
\nonumber \\
&&\times
\left[P_{qg}(-z_2)-P_{qg}(z_2)\right]
+{\cal O}(Q_\perp^2)
\nonumber \\
w^{qg}_{\Delta\Delta} 
&-& 
\frac{1}{2}\,
\frac{Q_\perp^2/Q^2}{1+Q_\perp^2/Q^2}\,
  \frac{w_{qg}^{\rm Asym}}{2}\
\Rightarrow \,
\frac{1}{2}\, e_q^2\,\frac{8\pi^2\alpha_s}{3x_1x_2}\,\delta(1-z_1)
\nonumber \\
&& \times
\left[P_{qg}(-z_2)-P_{qg}(z_2)\right]
+{\cal O}(Q_\perp^2) .
\label{sf-finite-qg-qt0}
\end{eqnarray}
With $z_1$ and $z_2$ interchanged, Eq.~(\ref{sf-finite-qg-qt0})  
is also true for the gluon-quark subprocess.
Other than the non-logarithmic finite piece 
(as $Q_{\perp}/Q \rightarrow 0$) in the quark-gluon
contributions to $W_L$ and $W_{\Delta\Delta}$, 
the asymptotic tensor completely removes the leading term 
of the perturbatively calculated helicity structure functions 
as $Q_\perp/Q\to 0$.  

  
The parton-level asymptotic terms for the polarized quark-antiquark, 
quark-gluon, and gluon-quark subprocesses are  
\begin{eqnarray}
\frac{\Delta w_{q\bar{q}}^{\rm Asym}}{2}
&\approx & 
\frac{w_{q\bar{q}}^{\rm Asym}}{2}
\nonumber\\
\frac{\Delta w_{qg}^{\rm Asym}}{2} 
&\approx & 
e_q^2\,\frac{8\pi^2\alpha_s}{3x_1x_2}\,\frac{Q^2}{Q_\perp^2}\,
\Delta P_{qg}(z_2)\,\delta(1-z_1)\, ,
\nonumber\\
\frac{\Delta w_{gq}^{\rm Asym}}{2} 
&\approx & 
e_q^2\,\frac{8\pi^2\alpha_s}{3x_1x_2}\,\frac{Q^2}{Q_\perp^2}\,
\Delta P_{qg}(z_1)\,\delta(1-z_2)\, .
\label{w-asym-pol}
\end{eqnarray}
Since $\Delta\omega_{q\bar{q}}^{\mu\nu}=\omega_{q\bar{q}}^{\mu\nu}$,
and $\Delta w_{q\bar{q}}^{\rm Asym}=w_{q\bar{q}}^{\rm Asym}$, 
Eq.~(\ref{sf-finite-qq-qt0}) is true also for the polarized
quark-antiquark subprocess. 
 
The finite contributions in the parton-level helicity structure 
functions for polarized quark-gluon or gluon-quark subprocesses are 
not the same as those for the corresponding unpolarized subprocesses.
We find
\begin{eqnarray}
\Delta w^{qg}_T 
&-& 
\left(1-\frac{1}{2} \frac{Q_\perp^2/Q^2}{1+Q_\perp^2/Q^2} \right)
\frac{\Delta w_{qg}^{\rm Asym}}{2}
\nonumber\\
&\Rightarrow &
{\cal O}(Q_\perp^0)
\nonumber \\
\Delta w^{qg}_L 
&-&
\frac{Q_\perp^2/Q^2}{1+Q_\perp^2/Q^2}\,
  \frac{\Delta w_{qg}^{\rm Asym}}{2}
\nonumber\\
&\Rightarrow &
e_q^2\,\frac{8\pi^2\alpha_s}{3x_1x_2}\, \delta(1-z_1)
\nonumber \\
&& \times
\left[\Delta P_{qg}(-z_2)-\Delta P_{qg}(z_2)\right]
+{\cal O}(Q_\perp^2)
\nonumber \\
\Delta w^{qg}_{\Delta\Delta} 
&-& \frac{1}{2}\,
\frac{Q_\perp^2/Q^2}{1+Q_\perp^2/Q^2}\,
    \frac{\Delta w_{qg}^{\rm Asym}}{2}
\nonumber\\
&\Rightarrow &
\frac{1}{2}\, e_q^2\,\frac{8\pi^2\alpha_s}{3x_1x_2}\,\delta(1-z_1) 
\nonumber \\
&& \times
\left[\Delta P_{qg}(-z_2)-\Delta P_{qg}(z_2)\right]
+{\cal O}(Q_\perp^2) .
\label{sf-finite-dqg-qt0}
\end{eqnarray}
With $z_1$ and $z_2$ interchanged, Eq.~(\ref{sf-finite-dqg-qt0}) 
is also true for gluon-quark subprocess.

The uncanceled finite term in the helicity structure functions 
$W_L$ and $W_{\Delta\Delta}$ is proportional to
\begin{equation}
P_{qg}(-z_2)-P_{qg}(z_2) = 4 z_2 T_R\, , 
\label{pqgavg}
\end{equation}
for unpolarized initial partonic states, and to  
\begin{equation}
\Delta P_{qg}(-z_2)-\Delta P_{qg}(z_2) = - 4 z_2 T_R\, ,
\label{pqgdep}
\end{equation}
for the polarized initial partonic states.
Therefore, for the scattering of two polarized hadrons 
with the same helicity (both positive or negative), 
the quark-gluon contribution to the perturbatively finite term 
of the helicity structure functions $W_L$ and $W_{\Delta\Delta}$ 
vanishes as $Q_\perp/Q\rightarrow 0$.  This result is obtained because 
the perturbative contribution to the longitudinal and double-spin
flip helicity structure functions is proportional to 
$P_{qg}(-z_2) + \Delta P_{qg}(-z_2)$ in the limit of
$Q_\perp/Q\rightarrow 0$, the corresponding asymptotic 
term is proportional to $P_{qg}(z_2) + \Delta P_{qg}(z_2)$, 
and the difference vanishes due to Eqs.~(\ref{pqgavg}) 
and (\ref{pqgdep}).
We also observe that, at this order, 
the uncanceled term in the quark-gluon subprocess 
is proportional to the helicity flipping splitting function, 
\begin{equation}
P_{q^-g^+}(z)=P_{q^+g^-}(z) = T_R (1-z)^2\, .
\end{equation}
The finite term as $Q_{\perp}/Q \rightarrow 0$ 
for the quark-antiquark subprocess at 
this order is removed completely by the asymptotic term 
since the helicity flipping 
splitting function for the quark vanishes at this order, 
$P_{q^-q^+}(z)=P_{q^+q^-}(z) = 0$.

Our observations allow us to claim that transverse momentum
dependent factorization for the full hadronic tensor, which is the
basis for the Collins-Soper-Sterman $b$-space resummation, 
breaks at subleading power in the $Q_\perp/Q$ expansion,   
but only in the helicity flipping channel.  The breaking seems   
not to supply leading logarithmic terms.

The asymptotic current-conserving tensor introduced in last section 
is sufficient to remove all leading divergent terms in the 
perturbatively calculated hadronic tensor.  The 
logarithmic terms in the perturbatively calculated helicity structure
functions, $W_T$, $W_L$ and $W_{\Delta\Delta}$, are shown here to 
have the same origin as those in the angular-integrated cross section.
Therefore, for these helicity structure functions 
we can obtain a perturbatively finite difference as
\begin{equation}
W_i^{\rm Finite} \equiv
W_i^{\rm Pert} - W_i^{\rm Asym}\, ,
\end{equation}
with $i=T,L,\Delta\Delta$.

\section{Full hadronic tensor including transverse momentum resummation}

In this section we present expressions for the transverse momentum 
dependence of the structure functions incorporating resummation to all 
orders in $\alpha_s$ of the singular divergent behavior as 
$Q_{\perp}/Q \rightarrow 0$ and including the contributions at 
order $\alpha_s$ that are finite in the small $Q_{\perp}$ limit.  We 
begin first with a brief summary of the resummation formalism developed 
for the angular-integrated cross section. 
 
As explained above, when $Q_\perp \ll Q$, the $Q_\perp$ 
distribution of the helicity structure functions calculated
in conventional fixed-order perturbation theory
receives a large logarithmic term, $\ln(Q/Q_\perp)$, at every power of
$\alpha_s$, which is a direct consequence of the emission of 
soft and collinear gluons from the incident partons.  
Therefore, when $Q_\perp/Q$ is sufficiently small, the
convergence of the conventional perturbative expansion in powers of
$\alpha_s$ is impaired, and the logarithmic terms must be resummed.

Resummation of the large logarithmic terms can be carried out either in 
$Q_\perp$-space directly, or in the impact parameter, $b$-space, 
which is the Fourier conjugate of $Q_\perp$-space.  
It was first shown by
Dokshitzer, Diakonov, and Troian that in the double leading logarithm 
approximation, the dominant contributions in the small $Q_T$ 
region can be resummed into a Sudakov form factor \cite{DDT-qt}.  
By imposing transverse momentum conservation without assuming 
strong ordering in the transverse momenta of radiated gluons, 
Parisi and Petronzio introduced a $b$-space resummation method which 
allows one to resum some subleading logarithmic terms \cite{pp-b}.  
Using a renormalization group equation technique, 
Collins and Soper improved $b$-space resummation to resum 
all terms as singular as $\ln^m(Q^2/Q_\perp^2)/Q_\perp^2$, 
as $Q_\perp\rightarrow 0$ \cite{cs-b}.  Using this renormalization 
group improved $b$-space resummation, Collins, Soper, and 
Sterman (CSS) derived a formalism for the transverse momentum
distributions of vector boson production in hadronic collisions
\cite{css-resum}.  This CSS formalism, developed originally for 
angular-integrated vector boson production, casts the cross section 
in the following generic form \cite{css-resum}  
\begin{eqnarray}
\frac{d\sigma}{d^4 q} 
&=&
\frac{1}{(2\pi)^2}\int d^2b\, e^{i\vec{Q}_\perp \cdot \vec{b}}\,
\widetilde{W}(b,Q,x_1,x_2) 
\nonumber \\
&+& Y(Q_\perp,Q,x_1,x_2)\, .  
\label{css-gen}
\end{eqnarray}
The function $\widetilde{W}$ provides the dominant contribution 
when $Q_\perp\ll Q$, while the $Y$ term supplies contributions that 
are negligible for small $Q_\perp$ but become important in practice 
when $Q_\perp\sim Q$.  The function $\widetilde{W}$ in Eq.~(\ref{css-gen}) 
incorporates all powers of large logarithmic contributions from $\ln(1/b^2)$ 
to $\ln(Q^2)$.  It has the following form \cite{css-resum} 
\begin{equation}
\widetilde{W}(b,Q,x_1,x_2) = 
{\rm e}^{-S(b,Q)}\, \widetilde{W}(b,c/b,x_1,x_2)\, ,
\label{css-W-sol}
\end{equation}
where $c$ is a constant of order one \cite{css-resum}, and
\begin{equation}
S(b,Q) = \int_{c^2/b^2}^{Q^2}\, 
  \frac{d{\mu}^2}{{\mu}^2} \left[
  \ln\left(\frac{Q^2}{{\mu}^2}\right) 
  A(\alpha_s({\mu})) + B(\alpha_s({\mu})) \right] .
\end{equation}
Functions $A(\alpha_s)$ and $B(\alpha_s)$ may be calculated perturbatively  
in powers of $\alpha_s$ \cite{css-resum}.  
Function $\widetilde{W}(b,c/b,x_A,x_B)$ 
in Eq.~(\ref{css-W-sol}) depends only on one momentum scale, $1/b$, 
and it may be calculated perturbatively as long as $1/b$ is large
enough.  The large logarithms from $\ln(c^2/b^2)$ to $\ln(Q^2)$ 
in $\widetilde{W}(b,Q,x_1,x_2)$ are completely resummed into the 
exponential factor $\exp[-S(b,Q)]$.  
The finite $Y$ term is defined to be the 
difference between the cross section calculated in 
conventional fixed-order perturbation theory and the asymptotic 
cross section which is equal to the perturbative expansion of 
the resummed part of the cross section, 
the first term on the RHS of Eq.~(\ref{css-gen}).  

The function $\widetilde{W}(b,Q,x_1,x_2)$ of the CSS $b$-space
resummation formalism in Eq.~(\ref{css-gen}) is not exactly 
equal to the Fourier transform of the transverse momentum
distribution, but its Fourier transform reproduces all leading 
divergences of the type $\ln^m(Q^2/Q_\perp^2)/Q_\perp^2$ in the
perturbatively calculated transverse momentum spectrum 
when $Q_\perp/Q\to 0$.  Combined with the perturbatively  
finite $Y$ term, the Fourier transform of the resummed 
$\widetilde{W}(b,Q,x_1,x_2)$ gives a good description of 
heavy vector boson production at collider energies 
\cite{msu-resum,qz-resum}.

The transverse momentum dependence of the angular distribution of
leptons from the Drell-Yan mechanism is determined by the transverse 
momentum dependence of the helicity structure functions.  Only the 
transverse structure function $W_T$ has a leading divergence of 
the type $\ln^m(Q^2/Q_\perp^2)/Q_\perp^2$ as $Q_\perp/Q\to 0$.  It 
might be natural to consider the resummation of these large logarithms 
into $W_T$ \cite{Chiappetta:1986yg,bqy-wz,Ellis:1997sc}.  
However, as we demonstrate in
Eqs.~(\ref{w-asym}) and~(\ref{w-asym-cs}), electromagnetic 
current conservation requires that the leading 
logarithmic divergences of the structure functions $W_L$ and 
$W_{\Delta\Delta}$ share the same origin as those in
$W_T$ and those in the angular-integrated cross section.
All are included in one asymptotic function, $W^{\rm Asym}$.  
Resummation of the large logarithmic terms of the Drell-Yan 
helicity structure functions can therefore be accomplished in terms  
of the resummed contribution to the angular-integrated 
Drell-Yan cross section.
Referring to Eq.~(\ref{st-asym-cs}), we obtain the resummed
contribution to the helicity structure functions in the Collins-Soper
frame as 
\begin{eqnarray}
W_T^{\rm Resum}
&=& 
\left(1-\frac{1}{2} \frac{Q_\perp^2/Q^2}{1+Q_\perp^2/Q^2} \right)
\frac{W^{\rm Resum}}{2} \, ,
\nonumber \\
W_L^{\rm Resum}
&=& 
\frac{Q_\perp^2/Q^2}{1+Q_\perp^2/Q^2}\,
\frac{W^{\rm Resum}}{2}\, ,
\nonumber \\
W_{\Delta\Delta}^{\rm Resum}
&=& 
\frac{1}{2}\,\frac{Q_\perp^2/Q^2}{1+Q_\perp^2/Q^2}\,
\frac{W^{\rm Resum}}{2} \, .
\label{st-resum-cs}
\end{eqnarray}
All depend on the same QCD resummed expression $W^{\rm Resum}$ 
that pertains to the angular-integrated Drell-Yan cross 
section~\cite{css-resum}.  By comparing Eq.~(\ref{css-gen}) with 
Eq.~(\ref{x-sec-integ}), we obtain
\begin{equation}
\frac{\alpha_{\rm em}^2}{12\pi^3 S^2 Q^2}\,W^{\rm Resum} 
=
\frac{1}{(2\pi)^2}\int d^2b\, e^{i\vec{Q}_\perp \cdot \vec{b}}\,
\widetilde{W}(b,Q,x_1,x_2) .
\label{w-reum-b}
\end{equation}

In analogy to the CSS result for the angular-integrated cross 
section in Eq.~(\ref{css-gen}), the expressions for the full 
transverse momentum distribution of the helicity structure functions 
are 
\begin{eqnarray}
W_T 
&=& 
\left(1-\frac{1}{2} \frac{Q_\perp^2/Q^2}{1+Q_\perp^2/Q^2} \right)
\frac{W^{\rm Resum}}{2}
\nonumber \\
&& 
+\left[
W_T^{\rm Pert}-
\left(1-\frac{1}{2} \frac{Q_\perp^2/Q^2}{1+Q_\perp^2/Q^2} \right)
\frac{W^{\rm Asym}}{2}
\right]\, ,
\nonumber \\
W_L
&=& 
\frac{Q_\perp^2/Q^2}{1+Q_\perp^2/Q^2}\,
\frac{W^{\rm Resum}}{2}
\nonumber \\
&&
+\left[
W_L^{\rm Pert} -
\frac{Q_\perp^2/Q^2}{1+Q_\perp^2/Q^2}\,
\frac{W^{\rm Asym}}{2}
\right]\, ,
\nonumber \\
W_{\Delta\Delta}
&=& 
\frac{1}{2}\,\frac{Q_\perp^2/Q^2}{1+Q_\perp^2/Q^2}\,
\frac{W^{\rm Resum}}{2}
\nonumber \\
&&
+\left[
W_{\Delta\Delta}^{\rm Pert} -
\frac{1}{2}\,\frac{Q_\perp^2/Q^2}{1+Q_\perp^2/Q^2}\,
\frac{W^{\rm Asym}}{2}
\right]
\label{st-full-cs} 
\end{eqnarray}
with the asymptotic term in these expressions equal to the perturbative 
expansion of the resummed contribution in powers of $\alpha_s$.  
As in the case of the angular-integrated cross section, these 
expressions are applicable for $Q\gtrsim Q_{\perp}$.  
Other effects must be considered when 
$Q_{\perp}\gg Q$~\cite{Berger:1998ev,Qiu:2001ac,Berger:2001wr}.   

Substituting the expressions for the helicity
structure functions $W_T$ and $W_L$ from Eq.~(\ref{st-full-cs})  
into Eq.~(\ref{x-sec-integ}), we obtain
\begin{eqnarray}
\frac{d\sigma}{d^4q} 
&=& 
\frac{\alpha_{\rm em}^2}{12\pi^3 S^2 Q^2}
\bigg[W^{\rm Resum}
\nonumber \\
&& + \left(2 W_T^{\rm Pert} + W_L^{\rm Pert} \right) 
          - W^{\rm Asym} \bigg]\, .
\label{x-sec-integ-full}
\end{eqnarray}
Using Eq.~(\ref{css-gen}), we 
find that the perturbatively finite $Y$-term is 
\begin{eqnarray}
Y= \frac{\alpha_{\rm em}^2}{12\pi^3 S^2 Q^2}
   \bigg[\left(2 W_T^{\rm Pert} + W_L^{\rm Pert} \right) 
         - W^{\rm Asym}\bigg]\, .
\label{css-y}
\end{eqnarray}

\subsection{Lam-Tung relation}
The Lam-Tung relation states that the longitudinal and 
the double-spin-flip structure functions obey the equality 
$W_L=2W_{\Delta\Delta}$.  
Based on Eqs.~(\ref{st-asym-cs}) and (\ref{st-resum-cs}) and
the definition in Eq.~(\ref{st-full-cs}), we find that
possible violation of the relation can come only from the 
non-singular finite piece of the perturbative contribution.  
The resummed contribution is known to dominate the 
angular-integrated cross section in the region of small and 
modest $Q_{\perp}$, and, by extension, we expect it to dominate 
the behavior of $W_L$ and $W_{\Delta\Delta}$ in the same region.   
We conclude that violation of the Lam-Tung relation as a 
function of $Q_\perp$ should be relatively small, consistent 
with the results of perturbative calculations at order 
$\alpha_s^2$~\cite{nlo}, but demonstrated here to all orders 
in $\alpha_s$.  

An alternative way to state the Lam-Tung relation is in terms of 
the angular coefficients $\lambda$ and $\nu$, defined in 
Eq.~(\ref{angularpar}). It is expressed as $1-\lambda - 2\nu = 0$. 
We derive 
\begin{eqnarray}
\lambda 
&=& \frac{W_T-W_L}{W_T+W_L}
\approx \frac{W_T^{\rm Resum}-W_L^{\rm Resum}}
             {W_T^{\rm Resum}+W_L^{\rm Resum}}
\nonumber \\
&=& \frac{1-\frac{1}{2}Q_\perp^2/Q^2}
         {1+\frac{3}{2}Q_\perp^2/Q^2}\, ,
\nonumber \\
\nu 
&=& \frac{2W_{\Delta\Delta}}{W_T+W_L}
\approx \frac{2W_{\Delta\Delta}^{\rm Resum}}
             {W_T^{\rm Resum}+W_L^{\rm Resum}}
\nonumber \\
&=& \frac{Q_\perp^2/Q^2}
         {1+\frac{3}{2}Q_\perp^2/Q^2}\, .  
\label{angularpar-resum}
\end{eqnarray}
The analytic expressions in Eq.~(\ref{angularpar-resum}) were derived
first in Ref.~\cite{Collins:1978yt} based on the perturbative calculation 
of $q\bar{q}\to \gamma^* g$.  Our result is valid for all orders in 
$\alpha_s$ if we retain only the leading resummed contribution, and 
it is independent of the type of incident hadrons.

A recent analysis of Fermilab data shows 
reasonable agreement with the Lam-Tung relation for
moderate values of $Q_\perp$ \cite{Zhu:2006gx}, while
early data with pion beams show some violation
\cite{Falciano:1986wk,Guanziroli:1987rp,Conway:1989fs,Heinrich:1991zm}.

\subsection{Phenomenological example}

\begin{figure}[t!]
\begin{center} 
\psfig{file=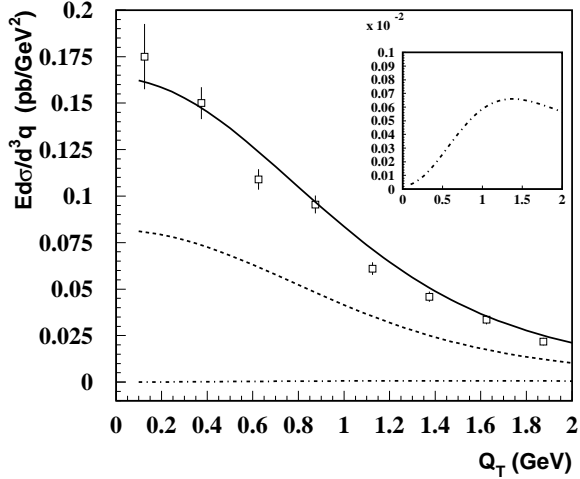,width=3.0in,angle=0}
\caption{The transverse momentum dependence 
of the angular-integrated Drell-Yan cross section, obtained from 
the contributions of the helicity structure functions, $W_T$ and 
$W_L$, in Eq.~(\protect\ref{st-full-cs}) is shown as a solid line 
and compared with data from Fermilab experiment E772 \protect\cite{E772} 
for $Q$ in the interval $(8,9)$~GeV.  The dashed and dot-dashed curves 
show our calculations for the contributions from $W_T$ and 
$W_L$. The inset shows the $W_L$ contribution on an expanded scale.    
}
\label{fig4:e772}
\end{center} 
\end{figure}

As an example, we show in Fig.~\ref{fig4:e772} an explicit numerical 
evaluation of the helicity structure functions $W_T$ and $W_L$ 
computed from Eq.~(\ref{st-full-cs}).  The double spin-flip structure 
function $W_{\Delta\Delta}=W_L/2$ since both the resummed contributions and 
the finite perturbative contributions at order $\alpha_s$ satisfy 
this relationship.  We choose the mass interval 8~GeV~$\leq Q \leq 9$~GeV and 
$E_{\rm beam}=800$~GeV in order to compare with data from 
Fermilab experiment E772~\cite{E772}. 
The parameters we use are identical to those used for Fig.~14 
in Ref.~\cite{qz-resum}.  

The dashed and dot-dashed lines in Fig.~\ref{fig4:e772} represent 
the $W_T$ and $W_L$ contributions to the cross section, while the
total contribution is proportional to $2W_T+W_L$.  We remark that the 
transverse momentum distribution after resummation is finite as 
$Q_\perp\to 0$ for $W_T$, but it becomes vanishingly small in the 
case of $W_L$.   

\subsection{Discussion of $W_{\Delta}$}

As shown in Sec.~III, the perturbative contribution to 
the single spin-flip structure 
function $W_\Delta$ is proportional to $Q/Q_\perp$, 
which is singular as $Q_\perp/Q\to 0$.  
Unlike the other helicity structure functions, 
$W_\Delta$ does not show a logarithmic 
divergence in the Collins-Soper frame, 
a feature that seems special for this frame \cite{bv-dy}. 
The absence of the divergence could be a consequence of the symmetry 
of the frame with respect to the hadron beam directions, which requires 
$W_\Delta \propto {\cal W}_1 {\rm e}^{-2y} - {\cal W}_2 {\rm e}^{2y}$ in 
Eq.~(\ref{bqr2cs}), and the fact that the leading logarithms arise 
from the region of phase space where $z_1\to 1$ and $z_2\to 1$.  

In this frame, the quark-antiquark contribution to 
$W_\Delta$ is completely antisymmetric in $z_1$ and $z_2$ 
because of the opposite sign between 
the ${\cal W}_1$ and the ${\cal W}_2$ terms above.  The quark-gluon 
(or gluon-quark) contribution is proportional to $1-z_2$ (or $(1-z_1)$). 
The asymmetry in $z_1$ and $z_2$ strongly reduces the numerical 
size of these contributions when $Q_\perp\neq 0$. 
 
The combination of the quark-antiquark 
and antiquark-quark subprocesses gives the 
following perturbative contribution to $W_\Delta$, 
\begin{equation}
W_{q\bar{q}}+W_{\bar{q}q} 
\propto 
\left[
 q_A(\xi_1)\, \bar{q}_B(\xi_2)
+\bar{q}_A(\xi_1)\, q_B(\xi_2)
\right]
\left(z_1^2 - z_2^2\right)\, .
\label{wd-qqb}
\end{equation}
This contribution vanishes in the central region 
for collisions between hadrons of the same type.  
The quark-gluon contribution also shows a similar asymmetry between
$z_1$ and $z_2$,
\begin{eqnarray}
W_{qg}+W_{gq} 
&\propto & 
 q_A(\xi_1)\, g_B(\xi_2)
\left(z_1^2 - 2z_2^2\right)
\nonumber \\
&&
+g_A(\xi_1)\, q_B(\xi_2)
\left(2z_1^2 - z_2^2\right)\, .
\label{wd-qg}
\end{eqnarray}

Collinear factorization in the perturbative calculation ceases to be 
valid when $Q_\perp \sim \Lambda_{\rm QCD}$ or less.  At $Q_\perp=0$, the 
helicity structure function $W_{\Delta}$ itself is ill-defined.  
We might still be able to test the physics of the single spin-flip 
structure function in the small $Q_{\perp}$ region by introducing a new 
observable, for example, the first moment of the structure function, 
\begin{equation}
\widetilde{W}_{\Delta}(Q_T,Q) 
\equiv
\int_0^{Q_T} dQ_{\perp} Q_{\perp} W_{\Delta}(Q_{\perp},Q)
\label{wd-1st}
\end{equation}
which is perturbatively more stable if $Q_T$ is large enough.

Is it possible that a different kind of resummation would handle the  
non-physical $Q_{\perp}^{-1}$ divergence at $Q_\perp=0$ in $W_\Delta$?  
We do not have an answer to this question in the collinear QCD 
factorization approach. However, we might gain insight by 
investigating the angular distribution from another perspective 
- starting with transverse momentum
dependent quark-antiquark annihilation \cite{bdw-dy,Ji:2004xq}. 

\section{Summary and discussion }

Massive virtual photons, the $W$ boson, and the $Z$ boson have
important decay modes into pairs of leptons.  The angular distribution
of these leptons, measured in the rest-frame of the parent states,
determines the alignment (polarization) of the massive vector boson
and, consequently, supplies more precise information on the production 
dynamics than is accessible from the angular-integrated rate alone.  
An understanding of the expected angular distribution is also important 
for estimating corrections associated with limited angular acceptance in 
typical experiments.  The changes expected in the angular distribution as 
a function of the transverse momentum $Q_{\perp}$ of the vector states is 
a topic of considerable interest, both for refined tests of QCD and to 
reduce systematic uncertainties on the determination of the W boson 
mass \cite{bqy-wz,Ellis:1997sc}.  

In this paper, we calculate the transverse momentum $Q_\perp$ dependence 
of the four helicity structure functions for the production of a massive 
pair of leptons with pair invariant mass $Q$. These structure functions 
determine the angular distribution of the leptons in the pair rest frame.  
We work within the QCD collinear factorization approach valid for 
$Q_\perp > \Lambda_{\rm QCD}$.  Our goal is the prediction of the full 
$Q_{\perp}$ dependence of the four structure functions, including  
the region of small and intermediate $Q_{\perp}$ where the cross section 
takes on its largest values.
  
As also noted by others, when calculated at fixed order in QCD perturbation 
theory, the structure functions show unphysical inverse-power $Q_\perp^{-n}$ 
($n = 1$ or $2$) or logarithmic $\ln (Q/Q_{\perp})$ 
divergences, or both, as $Q_{\perp} \rightarrow 0$.  For the 
angular-integrated cross section, $d\sigma/d^4q$, it is 
well established that similar unphysical divergences can be removed
after resummation of the $\ln^m(Q^2/Q_\perp^2)/Q_\perp^2$  
singular terms from initial-state gluon emission to all orders in $\alpha_s$ 
\cite{DDT-qt,pp-b,cs-b,css-resum}. 

We begin our analysis with the observation that the four helicity structure 
functions cannot be independent at $Q_\perp=0$.  The general 
tensor decomposition in the virtual photon rest frame in
Eq.~(\ref{W-q0sfs}) is ill-defined at $Q_\perp=0$.  Then, we employ 
electromagnetic current conservation to construct a new 
asymptotic hadronic tensor that has the right degrees of freedom as 
$Q_\perp\to 0$ and embodies the minimal divergent behavior present at 
fixed-order in QCD perturbation theory.  We find that the leading logarithmic 
behavior of three of the helicity structure functions, $W_T$, $W_L$, and 
$W_{\Delta\Delta}$, has a unique origin. Its origin is the same as that of 
the divergence in the angular-integrated cross section.  We are able, 
therefore, to reduce the problem of transverse momentum resummation for 
$W_T$, $W_L$, and $W_{\Delta \Delta}$ to the known solution of transverse 
momentum resummation for the angular-integrated cross section \cite{css-resum}. 
We prove that the small $Q_{\perp}$ logarithmic divergences in $W_T$, $W_L$, 
and $W_{\Delta \Delta}$ may be resummed to all orders in the strong coupling 
strength $\alpha_s$, yielding well behaved predictions for the $Q_{\perp}$ 
dependences that satisfy the expected kinematic constraints at small $Q_{\perp}$.  
The fourth structure function, $W_{\Delta}$, requires a different treatment, 
as discussed in Sec. V.C.   

The main results of our research include the fact that electromagnetic current 
conservation uniquely ties the perturbative divergences as 
$Q_\perp/Q\to 0$ of the otherwise independent helicity structure functions 
$W_T, W_L,$ and $W_{\Delta\Delta}$ to the divergence 
of the angular-integrated cross section.  Second, the perturbative divergence in the 
angular-integrated cross section is sufficient to remove all leading small $Q_{\perp}$
divergences of the individual helicity structure functions.  Third, transverse 
momentum resummation of the angular-integrated cross section determines the resummation 
of the large logarithmic terms of the helicity structure functions $W_T$, $W_L$, and 
$W_{\Delta\Delta}$.  Finally, the approximate Lam-Tung relation between the 
longitudinal and the double-spin-flip structure functions is an all-orders 
consequence of current conservation for the leading perturbatively divergent terms. 

In further work, we intend to examine the $Q_{\perp}$ dependence of 
$W$ and $Z$ boson production, where parity violating terms introduce
additional helicity structure functions.  Decay of these intermediate 
bosons into their dilepton channels supplies accurate measurements of the
masses of the bosons.  For $W$ production, more accurate 
predictions for the angular distribution of the single observed lepton should 
complement the missing energy technique and lead to an improved determination 
of the mass.  The mass of the $W$ boson provides an electroweak observable that 
bounds the mass of the Higgs boson within the framework of the standard model of particle
physics \cite{tevatron-w-higgs}. 

The use of current conservation to establish connections between the divergences 
of different helicity functions at $Q_\perp\to 0$ in the Drell-Yan process
may have immediate application for improving QCD resummation and predictions for particle 
production or other observables in semi-inclusive deep-inelastic scattering (SIDIS).
Unlike the Drell-Yan process, the lepton angles in SIDIS cannot be integrated over 
fully because the measurement of the DIS kinematic variables $x_B$ and $Q^2$  
requires specification of the production angle of the lepton in the final state.  Like 
the Drell-Yan cross section, the different helicity structure functions in SIDIS
have a $\ln^m(Q^2/q_\perp^2)$ perturbative divergence at small values of the 
particle transverse momentum $q_\perp$, defined in the frame where the vector
boson and the colliding hadron are aligned with each other.  All helicity structure 
functions contribute to particle production in SIDIS.  Only the leading singular 
$\ln^m(Q^2/q_\perp^2)/q_\perp^2$ logarithms are resummed in existing QCD calculations 
\cite{mos-sidis,Nadolsky:1999kb}.  Inclusion of the effects of resummation for the 
individual structure functions, as described in 
this paper, should lead to more accurate predictions 
for SIDIS observables, such as particle energy flow and rapidity dependence, that could be 
sensitive to the relative size of the different helicity structure functions.

\begin{acknowledgments}
E.L.B. is supported by the U.~S.~Department of Energy, Division of High
Energy Physics, under Contract No.\ DE-AC-02-06CH11357. 
J-W.Q. is supported in part by the U.~S.~Department of Energy under
Grant No.\ DE-FG02-87ER40371 and in part by the Argonne University of
Chicago Joint Theory Institute (JTI) Grant
03921-07-137.  R.A.R. is supported in part by the U.~S.~Department of 
Energy under Grant No.\ DE-FG02-87ER40371.  E.L.B. thanks the Aspen Center 
for Physics for hospitality while part of this research was being done.  
We are grateful to Daniel Boer, John T. Donohue, and Werner
Vogelsang for valuable communications.
\end{acknowledgments}

\appendix

\section{Drell-Yan cross section and angular distribution}
\label{appendix-A}

In this appendix, we summarize the basic formalism for 
calculating the cross section for dilepton production in 
the Drell-Yan model and the angular distribution of the leptons.
The expressions in this appendix also establish our notation. 

We consider the scattering of two hadrons of momentum $P_1$
and $P_2$, respectively, that produces a virtual photon of four-momentum
$q$, $A(P_1)+B(P_2)\rightarrow \gamma^*(q)+X$, that in turn decays into 
a pair of leptons of momentum $l$ and $\bar{l}$, as sketched in 
Fig.~\ref{fig1:dy}.  The cross section for this Drell-Yan production 
process can be expressed as
\begin{equation}
\frac{d\sigma}{d^4q d\Omega}
=
\frac{\alpha_{\rm em}^2}{2(2\pi)^4 S^2 Q^4}
L_{\mu\nu}\, W^{\mu\nu} .  
\label{x-sec}
\end{equation}
The leptonic tensor is 
\begin{equation}
L_{\mu\nu} = 2
\left[l_\mu \bar{l}_\nu + l_\nu \bar{l}_\mu -
      l\cdot \bar{l}\, g_{\mu\nu}
\right]\, ,
\label{L-tensor}
\end{equation}
and the hadronic tensor is defined as
\begin{eqnarray}
W_{\mu\nu} 
&=&
S \sum_{X} \langle P_1 P_2 | J_\mu^{\dagger}(0) |X\rangle
           \langle X| J_\nu(0) |P_1 P_2 \rangle
\nonumber\\
&& 
\times (2\pi)^4\, \delta^4(P_1+P_2-q-\sum_x(p_x))
\nonumber\\
&=&
S \int d^4z\, {\rm e}^{i q\cdot z}\,
\langle P_1 P_2 | J_{\mu}^{\dagger}(0)\, J_{\nu}(z) |P_1 P_2\rangle ,
\label{W-def}
\end{eqnarray}
where $J_\mu$ is the electromagnetic current.
Electromagnetic current conservation, $q^\mu\,W_{\mu\nu}=0$, 
and the fact that electromagnetic and strong interactions  
are invariant under the parity and time-reversal transformation,
allows us to express the Lorentz tensor, $W_{\mu\nu}$, in terms of 
four independent Lorentz scalar functions \cite{lt-dy}. 
We choose the following four frame-independent scalar functions,
\begin{eqnarray}
W^{\mu\nu}
&\equiv &
\widetilde{P}_1^\mu \widetilde{P}_1^\nu {\cal W}_1
+\widetilde{P}_2^\mu \widetilde{P}_2^\nu {\cal W}_2
\nonumber \\
&+&
\frac{1}{2} \left[
\widetilde{P}_1^\mu \widetilde{P}_2^\nu 
+\widetilde{P}_2^\mu \widetilde{P}_1^\nu 
\right]{\cal W}_3
- \widetilde{g}^{\mu\nu}\, {\cal W}_4 .  
\label{wi-indep}
\end{eqnarray}
The dimensionless current-conserving tensor and the vectors
are defined as 
\begin{eqnarray}
\widetilde{g}^{\mu\nu} 
&\equiv & 
g^{\mu\nu}- \frac{q^\mu q^\nu}{q^2} \, ,
\nonumber\\
\widetilde{P}_1^\mu 
&\equiv &
\widetilde{g}^{\mu\nu}\, P_{1\nu} /\sqrt{S} \, ,
\nonumber\\
\widetilde{P}_2^\mu 
&\equiv &
\widetilde{g}^{\mu\nu}\, P_{2\nu} /\sqrt{S} \, ,
\label{conserveV}
\end{eqnarray}
with $q_\mu \widetilde{g}^{\mu\nu}=0$.  Our choice of 
the four frame-independent scalar functions is slightly
different from that in Ref.~\cite{lt-dy}.  We find that
this choice is convenient for connecting to the parton-level 
perturbative calculation discussed below.

By contracting the leptonic tensor $L_{\mu\nu}$ and 
hadronic tensor $W_{\mu\nu}$ in Eq.~(\ref{x-sec}), we can express  
the Drell-Yan cross section in terms of the four 
scalar functions ${\cal W}_i$ and the measured hadron and lepton 
momenta.  

The physical meaning of the scalar functions can be 
appreciated if we express them in terms of the four independent 
``helicity'' structure functions, $W_{i}$ 
with $i=T,L,\Delta$, and $\Delta\Delta$, corresponding to the transverse
spin, longitudinal spin, single spin flip, and double spin flip
contributions to the Drell-Yan cross section \cite{lt-dy}.  
The helicity structure functions are defined in the dilepton 
center-of-mass frame (the virtual photon's rest frame).  

The full hadronic tensor in Eq.~(\ref{wi-indep}) can be also  
written in terms of the helicity structure functions and unit
vectors in the virtual photon rest frame as 
in Eq.~(\ref{W-q0sfs}) \cite{lt-dy}.
In this frame, the lepton momenta are 
\begin{eqnarray}
l^\mu &=&
\frac{Q}{2} \left(
1, \sin\theta \cos\phi, \sin\theta \sin\phi, \cos\theta \right) 
\nonumber\\
\bar{l}^\mu &=&
\frac{Q}{2} \left(
1, -\sin\theta \cos\phi, -\sin\theta \sin\phi, -\cos\theta 
\right)\, . 
\label{leptonM} 
\end{eqnarray}
Substituting the hadronic tensor in Eq.~(\ref{W-q0sfs}) and 
the leptonic tensor in Eq.~(\ref{leptonM}) into Eq.~(\ref{x-sec}),
one gets the differential cross section of Eq.~(\ref{x-sec-angular}). 

The frame-independent structure functions and the helicity
structure functions are uniquely related to each other once 
we make a choice of the coordinate system, or the unit vectors,
in the virtual photon rest frame.  
The unit vectors for the Collins-Soper frame are chosen as
\cite{cs-frame} 
\begin{eqnarray}
Z^\mu &=& 
\frac{2}{\sqrt{Q^2+Q_\perp^2}}
\left[ q_{P_2}\, \widetilde{P}_1^\mu
      -q_{P_1}\, \widetilde{P}_2^\mu  \right]\, ,
\nonumber \\
X^\mu &=& 
- \left(\frac{Q}{Q_\perp}\right) 
\frac{2}{\sqrt{Q^2+Q_\perp^2}}
\left[ q_{P_2}\, \widetilde{P}_1^\mu
      +q_{P_1}\, \widetilde{P}_2^\mu  \right]\,
\nonumber\\
Y^\mu &=& \epsilon^{\mu\nu\alpha\beta}\,T_\nu Z_\alpha X_\beta .
\label{cs-def0}
\end{eqnarray}
The dimensionless current-conserving hadron momenta, 
$\widetilde{P}_1^\mu$ and $\widetilde{P}_2^\mu$, 
are defined in Eq.~(\ref{conserveV}), and 
$q_{P_i} \equiv P_i\cdot q/\sqrt{S}$ with $i=1,2$.  
The hadron and the virtual photon momenta can be expressed in 
the center-of-mass frame of the collision as 
\begin{eqnarray}
P_1^\mu &=& 
\sqrt{\frac{S}{2}}\, \bar{n}^\mu\, ,
\quad\
P_2^\mu = \sqrt{\frac{S}{2}}\, n^\mu \, ,
\nonumber\\
q^\mu 
&=&
Q^+ \bar{n}^\mu + Q^- \bar{n}^\mu + Q_\perp n_\perp^\mu\, ,
\label{lab-mom}
\end{eqnarray}
with total center-of-mass collision energy $\sqrt{S}$, 
$Q^+=\sqrt{(Q^2+Q_\perp^2)/2}\, e^{y}$, and
$Q^-=\sqrt{(Q^2+Q_\perp^2)/2}\, e^{-y}$.  
In Eq.~(\ref{lab-mom}), $\bar{n}^\mu = \delta^{\mu +}$,
$n^\mu = \delta^{\mu -}$, and $n_\perp^\mu = \delta^{\mu\perp}$
are unit vectors that specify the light-cone coordinates of the 
collision center-of-mass frame, with $n^2 = \bar{n}^2 = 0$, 
$n^2_\perp = -1$, $n\cdot\bar{n} = 1$, and 
$n_\perp\cdot n = n_\perp\cdot\bar{n} = 0$.
In the Collins-Soper frame, the helicity structure functions 
can be expressed in terms of the frame-independent structure 
functions in Eq.~(\ref{wi-indep}) as 
\begin{eqnarray}
W_T
&=&
{\cal W}_4
+ \frac{1}{2}\frac{Q_\perp^2}{Q^2} 
\left[
\frac{1}{4}\left( {\cal W}_1\, {\rm e}^{-2y} 
                 +{\cal W}_2\, {\rm e}^{+2y} \right)
+\frac{1}{4} {\cal W}_3
\right],
\nonumber \\
W_L 
&=&
\frac{1}{4}\left( {\cal W}_1\, {\rm e}^{-2y} 
                 +{\cal W}_2\, {\rm e}^{+2y} \right)
-\frac{1}{4} {\cal W}_3
+ {\cal W}_4\, ,
\nonumber \\
W_{\Delta\Delta}
&=&
- \frac{1}{2}\frac{Q_\perp^2}{Q^2} 
\left[
\frac{1}{4}\left( {\cal W}_1\, {\rm e}^{-2y} 
                 +{\cal W}_2\, {\rm e}^{+2y} \right)
+\frac{1}{4} {\cal W}_3
\right],
\nonumber \\
W_{\Delta}
&=&
\frac{Q_\perp}{Q} 
\left[
 \frac{1}{4}{\cal W}_1\, {\rm e}^{-2y} 
-\frac{1}{4} {\cal W}_2\, {\rm e}^{+2y}
\right] \, .
\label{bqr2cs}
\end{eqnarray}
From the QCD collinear factorization formalism for the hadronic tensor
in Eq.~(\ref{Wmn-fac}) we obtain similar factorized relations for 
structure functions, 
\begin{equation}
W_i 
=
\sum_{ab}
\int \frac{d\xi_1}{\xi_1}  \int \frac{d\xi_2}{\xi_2}\,
\phi_a(\xi_1)\, \phi_b(\xi_2)\, w_i(\xi_1,\xi_2,q) ,
\label{W-fac}
\end{equation}
with $i=T,L,\Delta\Delta,\Delta$; and 
\begin{equation}
{\cal W}_i 
=
\sum_{ab}
\int \frac{d\xi_1}{\xi_1}  \int \frac{d\xi_2}{\xi_2}\,
\phi_a(\xi_1)\, \phi_b(\xi_2)\, \omega_i(\xi_1,\xi_2,q)
\label{calW-fac}
\end{equation}
with $i=1,2,3,4$.  
Using Eq.~(\ref{bqr2cs}), we derive the corresponding relation 
between the short-distance parton-level structure functions: 
\begin{eqnarray}
w_T
&=&
\omega_4
+ \frac{1}{2}\frac{Q_\perp^2}{Q^2} 
\left[
\frac{1}{4}\left( \omega_1\, {\rm e}^{-2y} 
                 +\omega_2\, {\rm e}^{+2y} \right)
+\frac{1}{4} \omega_3
\right],
\nonumber \\
w_L
&=&
\frac{1}{4}\left( \omega_1\, {\rm e}^{-2y} 
                 +\omega_2\, {\rm e}^{+2y} \right)
-\frac{1}{4} \omega_3
+ \omega_4 \, ,
\nonumber \\
w_{\Delta\Delta}
&=&
- \frac{1}{2}\frac{Q_\perp^2}{Q^2} 
\left[
\frac{1}{4}\left( \omega_1\, {\rm e}^{-2y} 
                 +\omega_2\, {\rm e}^{+2y} \right)
+\frac{1}{4} \omega_3
\right],
\nonumber \\
w_{\Delta}
&=&
\frac{Q_\perp}{Q} 
\left[
 \frac{1}{4}\,\omega_1\, {\rm e}^{-2y} 
-\frac{1}{4}\,\omega_2\, {\rm e}^{+2y}
\right]\, .
\label{bqr2cs-p}
\end{eqnarray}


Integration over the solid angle of the decay leptons gives 
the angular-integrated Drell-Yan cross section,
\begin{eqnarray}
\frac{d\sigma}{d^4q}
&=&
\frac{\alpha_{\rm em}^2}{12\pi^3 S^2 Q^2}\,
\left(2W_T+W_L\right)
\nonumber \\
&=&
\frac{\alpha_{\rm em}^2}{12\pi^3 S^2 Q^2}\,
\left(-g_{\mu\nu}W^{\mu\nu}\right)\, .
\label{x-sec-no-angular}
\end{eqnarray}
One can write the normalized Drell-Yan angular distribution as 
\begin{eqnarray}
\frac{dN}{d\Omega}
&\equiv &
\left( \frac{d\sigma}{d^4q} \right)^{-1}
\frac{d\sigma}{d^4q d\Omega}
\nonumber \\
&=&
\frac{3}{4\pi}\left(\frac{1}{\lambda +3}\right)
\bigg[ 1 + \lambda \cos^2\theta
\label{angulardist}\\
&&+ 
\mu \sin(2\theta)\cos\phi
+\frac{\nu}{2}\sin^2\theta\cos(2\phi) \bigg] ,
\nonumber 
\end{eqnarray}
with the coefficients of the angular dependence given by
\begin{eqnarray}
\lambda 
&=& \frac{W_T-W_L}{W_T+W_L}\, ,
\nonumber \\
\mu 
&=& \frac{W_{\Delta}}{W_T+W_L}\, ,
\nonumber \\
\nu 
&=& \frac{2W_{\Delta\Delta}}{W_T+W_L}.
\label{angularpar}
\end{eqnarray}

\section{Perturbative contributions from unpolarized partonic states}
\label{appendix-B}

In this appendix we summarize the perturbative contributions
to the parton-level helicity structure functions for unpolarized 
initial-state partons.

Using the definition in Eq.~(\ref{Wmn-fac}), we derive the
contribution to the parton-level hadronic tensor from the 
quark-antiquark annihilation diagrams in Fig.~\ref{fig2:qqb},  
with unpolarized initial parton states.  
\begin{eqnarray}
\omega_{q\bar{q}}^{\mu\nu}
&=&
\frac{4}{9}\,e_q^2\,\frac{8\pi^2\alpha_s}{\hat{t}\hat{u}}
\bigg[ 
-4\xi_1^2 Q^2\, S\, \widetilde{P}_1^\mu \widetilde{P}_1^\nu
-4\xi_2^2 Q^2\, S\, \widetilde{P}_2^\mu \widetilde{P}_2^\nu 
\nonumber\\
&& \hskip 0.8in
-((Q^2-\hat{t})^2 + (Q^2-\hat{u})^2 )\, \widetilde{g}^{\mu\nu}
\bigg]
\nonumber \\
&& \hskip 0.8in \times\,
S\,\delta(\hat{s}+\hat{t}+\hat{u}-Q^2)\, ,
\label{qqbar-tensor}
\end{eqnarray}
where $4/9=(1/3)^2\sum_A {\rm Tr}[t^A\, t^A]$ is the color factor
with SU(3) generator $t^A$, and $8\pi^2\alpha_s = (2\pi) g_s^2$.   
The factor $(2\pi)$ comes from the phase space expression 
\begin{eqnarray}
&&
S\,(2\pi)^4\,\delta^4(p_1+p_2-q-p_4)\,
\frac{d^3p_4}{(2\pi)^3 2E_4}
\nonumber\\
&& \hskip 0.2in 
= 2\pi\, S\,\delta(\hat{s}+\hat{t}+\hat{u}-Q^2)\, .
\label{ps-delta}
\end{eqnarray}
The parton-level Mandelstam variables are
\begin{eqnarray}
\hat{s}&=&(p_1+p_2)^2=\xi_1\, \xi_2\, S\, ,
\nonumber \\
\hat{t}&=&(p_1-q)^2=Q^2-2\xi_1 P_1\cdot q \, ,
\nonumber \\
\hat{u}&=&(p_2-q)^2=Q^2-2\xi_2 P_2\cdot q\, .
\label{stu-hat}
\end{eqnarray}
Using Eqs.~(\ref{qqbar-tensor}) and (\ref{bqr2cs-p}), we 
obtain the parton-level frame-independent structure functions 
\begin{eqnarray}
\omega^{q\bar{q}}_1 
&=& 
\frac{4}{9}\,e_q^2\,\frac{8\pi^2\alpha_s}{\hat{t}\hat{u}}\,
\left[ - 4 \xi_1^2\, Q^2\,S\,\right]\, 
S\,\delta(\hat{s}+\hat{t}+\hat{u}-Q^2)\, ,
\nonumber\\
\omega^{q\bar{q}}_2
&=& 
\frac{4}{9}\,e_q^2\,\frac{8\pi^2\alpha_s}{\hat{t}\hat{u}}\,
\left[ - 4 \xi_2^2\, Q^2\,S\,\right]\, 
S\,\delta(\hat{s}+\hat{t}+\hat{u}-Q^2)\, ,
\nonumber\\
\omega^{q\bar{q}}_3
&=& 
0
\nonumber\\
\omega^{q\bar{q}}_4
&=& 
\frac{4}{9}\,e_q^2\,\frac{8\pi^2\alpha_s}{\hat{t}\hat{u}}\,
\left[ \xi_1^2\, {\rm e}^{-2y} + \xi_2^2\, {\rm e}^{2y} \right]\, 
\left(Q^2+Q_\perp^2\right) S\,
\nonumber\\
&& \hskip 0.6in 
\times\,
S\,\delta(\hat{s}+\hat{t}+\hat{u}-Q^2)\, ;
\label{qqbar-sf}
\end{eqnarray}
and the corresponding parton-level helicity structure functions 
in the Collins-Soper frame,
\begin{eqnarray}
w^{q\bar{q}}_T 
&=& 
\frac{4}{9}\,e_q^2\,\frac{8\pi^2\alpha_s}{\hat{t}\hat{u}}\,
\left[ \xi_1^2\, {\rm e}^{-2y} + \xi_2^2\, {\rm e}^{2y} \right]\, 
S(Q^2+\frac{1}{2} Q_\perp^2)\, 
\nonumber \\
&& \hskip 0.4in
\times\,
S\,\delta(\hat{s}+\hat{t}+\hat{u}-Q^2)\, ,
\nonumber\\
w^{q\bar{q}}_L
&=& 
\frac{4}{9}\,e_q^2\,\frac{8\pi^2\alpha_s}{\hat{t}\hat{u}}\,
\left[ \xi_1^2\, {\rm e}^{-2y} + \xi_2^2\, {\rm e}^{2y} \right]\, 
\left(S\, Q_\perp^2\right)\, 
\nonumber \\
&& \hskip 0.4in
\times\,
S\,\delta(\hat{s}+\hat{t}+\hat{u}-Q^2)\, ,
\nonumber\\
w^{q\bar{q}}_{\Delta\Delta}
&=& 
\frac{4}{9}\,e_q^2\,\frac{8\pi^2\alpha_s}{\hat{t}\hat{u}}\,
\left[ \xi_1^2\, {\rm e}^{-2y} + \xi_2^2\, {\rm e}^{2y} \right]\, 
\left(\frac{1}{2}\,S\, Q_\perp^2\right)\, 
\nonumber \\
&& \hskip 0.4in
\times\,
S\,\delta(\hat{s}+\hat{t}+\hat{u}-Q^2)
\nonumber\\
&=& \frac{1}{2}\, w^{q\bar{q}}_L\,\, ,
\nonumber\\
w^{q\bar{q}}_\Delta
&=& 
\frac{4}{9}\,e_q^2\,\frac{8\pi^2\alpha_s}{\hat{t}\hat{u}}\,
\left[ -\xi_1^2\, {\rm e}^{-2y} + \xi_2^2\, {\rm e}^{2y} \right]\, 
\left(S\,Q^2\right)\frac{Q_\perp}{Q}\, 
\nonumber \\
&& \hskip 0.4in
\times\,
S\,\delta(\hat{s}+\hat{t}+\hat{u}-Q^2)\, .
\label{qqbar-sf-cs}
\end{eqnarray}

From the quark-gluon scattering diagrams in Fig.~\ref{fig3:qg}
with unpolarized initial parton states, we derive the quark-gluon 
contribution to the parton-level hadronic tensor  
\begin{eqnarray}
\omega_{qg}^{\mu\nu}
&=&
\frac{1}{6}\,e_q^2\,\frac{8\pi^2\alpha_s}{\hat{s}(-\hat{t})}
\bigg[ 
-8\xi_1^2\, Q^2\, S\, \widetilde{P}_1^\mu \widetilde{P}_1^\nu
-4\xi_2^2\, Q^2\, S\, \widetilde{P}_2^\mu \widetilde{P}_2^\nu 
\nonumber\\
&& \hskip 0.75in
-4\xi_1\,\xi_2\, Q^2\, S\, 
  [ \widetilde{P}_1^\mu \widetilde{P}_2^\nu
   +\widetilde{P}_2^\mu \widetilde{P}_1^\nu ]
\nonumber \\
&& \hskip 0.75in
-((Q^2-\hat{t})^2 + (Q^2-\hat{s})^2 )\, \widetilde{g}^{\mu\nu}
\bigg]\,
\nonumber\\
&& \hskip 0.75in 
\times \,
S\,\delta(\hat{s}+\hat{t}+\hat{u}-Q^2) ,
\label{qg-tensor}
\end{eqnarray}
where $1/6=(1/3)(1/8)\sum_A {\rm Tr}[t^A\, t^A]$ is the color 
factor.  We obtain the parton-level frame-independent structure 
functions 
\begin{eqnarray}
\omega^{qg}_1 
&=& 
\frac{1}{6}\,e_q^2\,\frac{8\pi^2\alpha_s}{\hat{s}(-\hat{t})}\,
\left[ - 8 \xi_1^2\, Q^2\,S\,\right]\, 
S\,\delta(\hat{s}+\hat{t}+\hat{u}-Q^2)\, ,
\nonumber\\
\omega^{qg}_2
&=& 
\frac{1}{6}\,e_q^2\,\frac{8\pi^2\alpha_s}{\hat{s}(-\hat{t})}\,
\left[ - 4 \xi_2^2\, Q^2\,S\,\right]\, 
S\,\delta(\hat{s}+\hat{t}+\hat{u}-Q^2)\, ,
\nonumber\\
\omega^{qg}_3
&=& 
\frac{1}{6}\,e_q^2\,\frac{8\pi^2\alpha_s}{\hat{s}(-\hat{t})}\,
\left[ - 8 \xi_1\, \xi_2\, Q^2\,S\,\right]\, 
S\,\delta(\hat{s}+\hat{t}+\hat{u}-Q^2)\, ,
\nonumber\\
\omega^{qg}_4
&=& 
\frac{1}{6}\,e_q^2\,\frac{8\pi^2\alpha_s}{\hat{s}(-\hat{t})}\,
\bigg[ \xi_1^2\,{\rm e}^{-2y}\, S(Q^2+Q_\perp^2)
\nonumber\\
&& \hskip 0.3in
+(Q^2-\xi_1\xi_2 S)^2 \bigg]\, 
S\,\delta(\hat{s}+\hat{t}+\hat{u}-Q^2)\, ;
\label{qg-sf}
\end{eqnarray}
and the corresponding contribution to the parton-level helicity structure
functions in the Collins-Soper frame,  
\begin{eqnarray}
w^{qg}_T 
&=& 
\frac{1}{6}\,e_q^2\,\frac{8\pi^2\alpha_s}{\hat{s}(-\hat{t})}\,
\bigg[ \xi_1^2\, {\rm e}^{-2y}\,S Q^2 
      + (Q^2-\xi_1\xi_2 S)^2
\nonumber\\
&& \hskip 0.7in 
-\frac{1}{2}\frac{Q^2_\perp}{Q^2}
\left[\xi_2^2\, {\rm e}^{2y} + 2\xi_1\xi_2 \right] S Q^2
\bigg]\,
\nonumber \\
&& \hskip 0.7in 
\times
S\,\delta(\hat{s}+\hat{t}+\hat{u}-Q^2)\, ,
\nonumber\\
w^{qg}_L
&=& 
\frac{1}{6}\,e_q^2\,\frac{8\pi^2\alpha_s}{\hat{s}(-\hat{t})}\,
\left[ 2 \xi_1^2\, {\rm e}^{-2y} + \xi_2^2\, {\rm e}^{2y} 
      + 2 \xi_1 \xi_2 
\right]\, 
\nonumber \\
&& \hskip 0.6in 
\times
\left(S\, Q_\perp^2\right) \, 
S\,\delta(\hat{s}+\hat{t}+\hat{u}-Q^2)\, ,
\nonumber\\
w^{qg}_{\Delta\Delta}
&=& 
\frac{1}{6}\,e_q^2\,\frac{8\pi^2\alpha_s}{\hat{s}(-\hat{t})}\,
\left[ 2 \xi_1^2\, {\rm e}^{-2y} + \xi_2^2\, {\rm e}^{2y} 
      + 2 \xi_1 \xi_2 \right]\, 
\nonumber \\
&& \hskip 0.6in 
\times
\left(\frac{1}{2}\,S\, Q_\perp^2\right)
S\,\delta(\hat{s}+\hat{t}+\hat{u}-Q^2)
\nonumber\\
&=&\frac{1}{2}\,
w^{qg}_L
\nonumber\\
w^{qg}_\Delta
&=& 
\frac{1}{6}\,e_q^2\,\frac{8\pi^2\alpha_s}{\hat{s}(-\hat{t})}\,
\left[ -2 \xi_1^2\, {\rm e}^{-2y} + \xi_2^2\, {\rm e}^{2y} \right]
\left(\frac{Q_\perp}{Q}\right) 
\nonumber \\
&& \hskip 0.6in 
\times
\left(S\,Q^2\right)
S\,\delta(\hat{s}+\hat{t}+\hat{u}-Q^2)\, .
\label{qg-sf-cs}
\end{eqnarray}

Similarly, we derive the contributions to the parton-level hadronic tensor
from the gluon-quark scattering diagrams.  They are the same as those 
from the quark-gluon scattering diagrams with the momenta $p_1$ and $p_2$ 
(or equivalently with $\hat{t}$ and $\hat{u}$, and $\xi_1$ and $\xi_2$) 
interchanged.

\section{Perturbative contributions from polarized partonic states}
\label{appendix-C}

In this appendix we summarize the perturbative contributions
to the parton-level helicity structure functions for 
polarized initial-state partons, defined as the states with
incoming parton polarization projected onto the {\it difference} of
the parton helicity states.  

Based on the same quark-antiquark annihilation diagrams 
in Fig.~\ref{fig2:qqb}, we find at this order that the contribution to
the parton-level hadronic tensor from the scattering of a polarized
incoming quark and antiquark is the same as that from the 
scattering of an unpolarized quark and antiquark,  
\begin{equation}
\Delta \omega_{q\bar{q}}^{\mu\nu}
=\omega_{q\bar{q}}^{\mu\nu} \, .
\label{qqb-pol}
\end{equation}

On the other hand, the quark-gluon scattering diagrams in
Fig.~\ref{fig3:qg} with polarized quark and gluon
initial states give a contribution to the parton-level hadronic 
tensor that differs from that for scattering of an unpolarized
quark and gluon,
\begin{equation}
\Delta\omega_{qg}^{\mu\nu}
\neq \omega_{qg}^{\mu\nu} \, .
\label{qg-pol}
\end{equation}
We derive 
\begin{eqnarray}
\Delta\omega_{qg}^{\mu\nu}
&=&
\frac{1}{6}\,e_q^2\,\frac{8\pi^2\alpha_s}{\hat{s}(-\hat{t})}
\bigg[ 
+4\xi_2^2\, Q^2\, S\, \widetilde{P}_2^\mu \widetilde{P}_2^\nu 
\nonumber \\
&& \hskip 0.75in
+4\xi_1\,\xi_2\, Q^2\, S\, 
  [ \widetilde{P}_1^\mu \widetilde{P}_2^\nu
   +\widetilde{P}_2^\mu \widetilde{P}_1^\nu ]
\nonumber \\
&& \hskip 0.75in
-((Q^2-\hat{t})^2 - (Q^2-\hat{s})^2 )\, 
\widetilde{g}^{\mu\nu}
\bigg]
\nonumber\\
&& \hskip 0.75in
\times\,
S\, \delta(\hat{s}+\hat{t}+\hat{u}-Q^2)\, .
\label{dqg-tensor}
\end{eqnarray}
The contributions to the parton-level frame-independent
structure functions are  
\begin{eqnarray}
\Delta\omega^{qg}_1 
&=& 
0\, ,
\nonumber\\
\Delta\omega^{qg}_2
&=& 
\frac{1}{6}\,e_q^2\,\frac{8\pi^2\alpha_s}{\hat{s}(-\hat{t})}\,
\left[ 4 \xi_2^2\, Q^2\,S\,\right] 
S\,\delta(\hat{s}+\hat{t}+\hat{u}-Q^2)\, ,
\nonumber\\
\Delta\omega^{qg}_3
&=& 
\frac{1}{6}\,e_q^2\,\frac{8\pi^2\alpha_s}{\hat{s}(-\hat{t})}\,
\left[ 8 \xi_1\, \xi_2\, Q^2\,S\,\right] 
S\,\delta(\hat{s}+\hat{t}+\hat{u}-Q^2)\, ,
\nonumber\\
\Delta\omega^{qg}_4
&=& 
\frac{1}{6}\,e_q^2\,\frac{8\pi^2\alpha_s}{\hat{s}(-\hat{t})}\,
\bigg[ \xi_1^2\, {\rm e}^{-2y} \, S(Q^2+Q_\perp^2)
\nonumber\\
&&      
-(Q^2-\xi_1\xi_2 S)^2 
\bigg]\, 
S\,\delta(\hat{s}+\hat{t}+\hat{u}-Q^2)\, .
\label{dqg-sf}
\end{eqnarray}
The corresponding contributions to the parton-level helicity structure
functions in the Collins-Soper frame are 
\begin{eqnarray}
\Delta w^{qg}_T 
&=& 
\frac{1}{6}\,e_q^2\,\frac{8\pi^2\alpha_s}{\hat{s}(-\hat{t})}\,
\bigg[ \xi_1^2\, {\rm e}^{-2y}\,S Q^2 
      - (Q^2-\xi_1\xi_2 S)^2 
\nonumber\\
&& \hskip 0.2in 
+\frac{1}{2}\frac{Q^2_\perp}{Q^2}
\left[ 2 \xi_1^2\, {\rm e}^{-2y} + \xi_2^2\, {\rm e}^{2y} 
      + 2\xi_1\xi_2 \right] S Q^2
\bigg]\,
\nonumber\\
&& \hskip 0.2in 
\times\,
S\,\delta(\hat{s}+\hat{t}+\hat{u}-Q^2)\, ,
\nonumber\\
\Delta w^{qg}_L
&=& 
\frac{1}{6}\,e_q^2\,\frac{8\pi^2\alpha_s}{\hat{s}(-\hat{t})}\,
\left[ - \xi_2^2\, {\rm e}^{2y} 
       - 2 \xi_1 \xi_2 
\right]\, 
\nonumber\\
&& \hskip 0.2in 
\times
\left(S\, Q_\perp^2\right)
S\,\delta(\hat{s}+\hat{t}+\hat{u}-Q^2)\, ,
\nonumber\\
\Delta w^{qg}_{\Delta\Delta}
&=& 
\frac{1}{6}\,e_q^2\,\frac{8\pi^2\alpha_s}{\hat{s}(-\hat{t})}\,
\left[ - \xi_2^2\, {\rm e}^{2y} 
       - 2 \xi_1 \xi_2 \right]\, 
\nonumber\\
&& \hskip 0.2in 
\times
\left(\frac{1}{2}\,S\, Q_\perp^2\right)
S\,\delta(\hat{s}+\hat{t}+\hat{u}-Q^2)
\nonumber\\
&=& 
\frac{1}{2}\,
\Delta w^{qg}_L
\nonumber\\
\Delta w^{qg}_\Delta
&=& 
\frac{1}{6}\,e_q^2\,\frac{8\pi^2\alpha_s}{\hat{s}(-\hat{t})}\,
\left[ - \xi_2^2\, {\rm e}^{2y} \right] 
\left(\frac{Q_\perp}{Q}\right) 
\nonumber\\
&& \hskip 0.2in 
\times
\left(S\,Q^2\right)
S\,\delta(\hat{s}+\hat{t}+\hat{u}-Q^2)\, .
\label{dqg-sf-cs}
\end{eqnarray}

Similarly, we derive the contribution from the polarized gluon and
quark scattering process,
\begin{eqnarray}
\Delta\omega_{gq}^{\mu\nu}
&=&
\frac{1}{6}\,e_q^2\,\frac{8\pi^2\alpha_s}{\hat{s}(-\hat{u})}
\bigg[ 
+4\xi_1^2\, Q^2\, S\, \widetilde{P}_1^\mu \widetilde{P}_1^\nu 
\nonumber \\
&& \hskip 0.8in
+4\xi_1\,\xi_2\, Q^2\, S\, 
  [ \widetilde{P}_1^\mu \widetilde{P}_2^\nu
   +\widetilde{P}_2^\mu \widetilde{P}_1^\nu ]
\nonumber \\
&& \hskip 0.8in
-((Q^2-\hat{u})^2 - (Q^2-\hat{s})^2 )\, 
\widetilde{g}^{\mu\nu}
\bigg]
\nonumber\\
&& \hskip 0.8in
\times\,
S\, \delta(\hat{s}+\hat{t}+\hat{u}-Q^2)\, .
\label{dgq-tensor}
\end{eqnarray}
The contributions to the parton-level frame-independent
structure functions are 
\begin{eqnarray}
\Delta\omega^{gq}_1 
&=& 
\frac{1}{6}\,e_q^2\,\frac{8\pi^2\alpha_s}{\hat{s}(-\hat{u})}\,
\left[ 4 \xi_1^2\, Q^2\,S\,\right] 
S\,\delta(\hat{s}+\hat{t}+\hat{u}-Q^2)\, ,
\nonumber\\
\Delta\omega^{gq}_2
&=& 
0\, ,
\nonumber\\
\Delta\omega^{gq}_3
&=& 
\frac{1}{6}\,e_q^2\,\frac{8\pi^2\alpha_s}{\hat{s}(-\hat{u})}\,
\left[ 8 \xi_1\, \xi_2\, Q^2\,S\,\right] 
S\,\delta(\hat{s}+\hat{t}+\hat{u}-Q^2)\, ,
\nonumber\\
\Delta\omega^{gq}_4
&=& 
\frac{1}{6}\,e_q^2\,\frac{8\pi^2\alpha_s}{\hat{s}(-\hat{u})}\,
\bigg[ \xi_2^2\, {\rm e}^{2y} \, S(Q^2+Q_\perp^2)
\nonumber\\
&&      
-(Q^2-\xi_1\xi_2 S)^2 
\bigg]\, 
S\,\delta(\hat{s}+\hat{t}+\hat{u}-Q^2)\, .
\label{dgq-sf}
\end{eqnarray}
The corresponding contributions to the parton-level helicity structure
functions in the Collins-Soper frame are 
\begin{eqnarray}
\Delta w^{gq}_T 
&=& 
\frac{1}{6}\,e_q^2\,\frac{8\pi^2\alpha_s}{\hat{s}(-\hat{u})}\,
\bigg[ \xi_2^2\, {\rm e}^{2y}\,S Q^2 
      - (Q^2-\xi_1\xi_2 S)^2 
\nonumber\\
&& \hskip 0.2in 
+\frac{1}{2}\frac{Q^2_\perp}{Q^2}
\left[ \xi_1^2\, {\rm e}^{-2y} + 2\xi_2^2\, {\rm e}^{2y} 
      + 2\xi_1\xi_2 \right] S Q^2
\bigg]\,
\nonumber\\
&& \hskip 0.2in 
\times\,
S\,\delta(\hat{s}+\hat{t}+\hat{u}-Q^2)\, ,
\nonumber\\
\Delta w^{gq}_L
&=& 
\frac{1}{6}\,e_q^2\,\frac{8\pi^2\alpha_s}{\hat{s}(-\hat{u})}\,
\left[ - \xi_1^2\, {\rm e}^{-2y} 
       - 2 \xi_1 \xi_2 
\right]\, 
\nonumber\\
&& \hskip 0.2in 
\times
\left(S\, Q_\perp^2\right)
S\,\delta(\hat{s}+\hat{t}+\hat{u}-Q^2)\, ,
\nonumber\\
\Delta w^{gq}_{\Delta\Delta}
&=& 
\frac{1}{6}\,e_q^2\,\frac{8\pi^2\alpha_s}{\hat{s}(-\hat{u})}\,
\left[ - \xi_1^2\, {\rm e}^{-2y} 
       - 2 \xi_1 \xi_2 \right]\, 
\nonumber\\
&& \hskip 0.2in 
\times
\left(\frac{1}{2}\,S\, Q_\perp^2\right)
S\,\delta(\hat{s}+\hat{t}+\hat{u}-Q^2)
\nonumber\\
&=& 
\frac{1}{2}\,
\Delta w^{gq}_L
\nonumber\\
\Delta w^{gq}_\Delta
&=& 
\frac{1}{6}\,e_q^2\,\frac{8\pi^2\alpha_s}{\hat{s}(-\hat{u})}\,
\left[ \xi_1^2\, {\rm e}^{-2y} \right] 
\left(\frac{Q_\perp}{Q}\right) 
\nonumber\\
&& \hskip 0.2in 
\times
\left(S\,Q^2\right)
S\,\delta(\hat{s}+\hat{t}+\hat{u}-Q^2)\, .
\label{dgq-sf-cs}
\end{eqnarray}



\begin{thebibliography}{999}

\bibitem{Drell:1970wh}
  S.~D.~Drell and T.~M.~Yan,
  Phys.\ Rev.\ Lett.\  {\bf 25}, 316 (1970)
  [Erratum-ibid.\  {\bf 25}, 902 (1970)].

\bibitem{earlyangulardata}  See, for example, 
G. E. Hogan {\em et al}, Phys. Rev. Lett. {\bf 42}, 948 (1979);
K. J. Anderson {\em et al}, Phys. Rev. Lett. {\bf 43}, 1219 (1979); 
J. Badier {\em et al} Z. Phys. {\bf C11}, 195 (1981) and
references therein.

\bibitem{bdw-dy}
  E.~L.~Berger, J.~T.~Donohue and S.~Wolfram,
  Phys.\ Rev.\  D {\bf 17}, 858 (1978);
  J.~T.~Donohue,
in {\it Phenomenology Of Quantum Chromodynamics, Proceedings of the 
XIII Rencontre de Moriond, 1978}, Edited by J. Tran Thanh Van (Editions 
Frontieres, Dreux, 1978), Vol.I, 159-163.

\bibitem{lt-dy}
  C.~S.~Lam and W.~K.~Tung,
  Phys.\ Rev.\  D {\bf 18}, 2447 (1978).

\bibitem{Kajantie:1978yp}
  K.~Kajantie, J.~Lindfors and R.~Raitio,
  Phys.\ Lett.\  B {\bf 74}, 384 (1978);
  J.~Lindfors,
  Phys.\ Scripta {\bf 20}, 19 (1979).

\bibitem{Cleymans:1978je}
  J.~Cleymans and M.~Kuroda,
  Phys.\ Lett.\  B {\bf 80}, 385 (1979)
  [Erratum-ibid.\  {\bf 86B}, 426 (1979)];
  Nucl.\ Phys.\  B {\bf 155}, 480 (1979)
  [Erratum-ibid.\  B {\bf 160}, 510 (1979)].

\bibitem{lt-dy2}
  C.~S.~Lam and W.~K.~Tung,
  Phys.\ Rev.\  D {\bf 21}, 2712 (1980).

\bibitem{Collins:1978yt}
  J.~C.~Collins,
  Phys.\ Rev.\ Lett.\  {\bf 42}, 291 (1979).

\bibitem{nlo}
  E.~Mirkes,
  Nucl.\ Phys.\  B {\bf 387}, 3 (1992);
  E.~Mirkes and J.~Ohnemus,
  Phys.\ Rev.\  D {\bf 51}, 4891 (1995).

\bibitem{Chiappetta:1986yg}
  P.~Chiappetta and M.~Le Bellac,
  Z.\ Phys.\  C {\bf 32}, 521 (1986).

\bibitem{bv-dy}
  D.~Boer and W.~Vogelsang,
  Phys.\ Rev.\  D {\bf 74}, 014004 (2006)
  [arXiv:hep-ph/0604177].

\bibitem{spin-whitepaper}
G.~Bunce, {\it et al.}, ``Status and Prospects of the RHIC
Spin Physics Program'', 
{\tt http://spin.rhic.bnl.gov/rsc/report/masterspin.pdf}, 
January, 2007.

\bibitem{bqy-wz}
  C.~Balazs, J.~W.~Qiu and C.~P.~Yuan,
  Phys.\ Lett.\  B {\bf 355}, 548 (1995)
  [arXiv:hep-ph/9505203].

\bibitem{Ellis:1997sc}
  R.~K.~Ellis, D.~A.~Ross and S.~Veseli,
  Nucl.\ Phys.\  B {\bf 503}, 309 (1997)
  [arXiv:hep-ph/9704239].

\bibitem{Falciano:1986wk}
  S.~Falciano {\it et al.}  [NA10 Collaboration],
  Z.\ Phys.\  C {\bf 31}, 513 (1986).

\bibitem{Guanziroli:1987rp}
  M.~Guanziroli {\it et al.}  [NA10 Collaboration],
  Z.\ Phys.\  C {\bf 37}, 545 (1988).

\bibitem{Conway:1989fs}
  J.~S.~Conway {\it et al.},
  Phys.\ Rev.\  D {\bf 39}, 92 (1989).

\bibitem{Heinrich:1991zm}
  J.~G.~Heinrich {\it et al.},
  Phys.\ Rev.\  D {\bf 44}, 1909 (1991).

\bibitem{Zhu:2006gx}
  L.~Y.~Zhu {\it et al.}  [FNAL-E866/NuSea Collaboration],
  Phys. Rev. Lett. {\bf 99}, 082301 (2007) [arXiv:hep-ex/0609005].

\bibitem{Berger:1979du}
  E.~L.~Berger and S.~J.~Brodsky,
  Phys.\ Rev.\ Lett.\  {\bf 42}, 940 (1979).

\bibitem{Qiu:1990xx}
  J.~W.~Qiu and G.~Sterman,
  Nucl.\ Phys.\  B {\bf 353}, 105 (1991).

\bibitem{DDT-qt}
  Y.~L.~Dokshitzer, D.~Diakonov and S.~I.~Troian,
  Phys.\ Rept.\  {\bf 58}, 269 (1980).

\bibitem{pp-b}
  G.~Parisi and R.~Petronzio,
  Nucl.\ Phys.\  B {\bf 154}, 427 (1979).

\bibitem{cs-b}
  J.~C.~Collins and D.~E.~Soper,
  Nucl.\ Phys.\  B {\bf 193}, 381 (1981)
  [Erratum-ibid.\  B {\bf 213}, 545 (1983)].

\bibitem{css-resum}
  J.~C.~Collins, D.~E.~Soper and G.~Sterman,
  Nucl.\ Phys.\  B {\bf 250}, 199 (1985), and references therein.

\bibitem{cs-frame}
  J.~C.~Collins and D.~E.~Soper,
  Phys.\ Rev.\  D {\bf 16}, 2219 (1977).

\bibitem{bqr-short}
  E.~L.~Berger, J.~W.~Qiu and R.~A.~Rodriguez-Pedraza,
  arXiv:0707.3150 [hep-ph], to be published in Physics Letters B.  

\bibitem{css-fac}
J.C.~Collins, D.E.~Soper, G.~Sterman,
Adv.\ Ser.\ Direct.\ High Energy Phys.\  {\bf 5}, 1 (1988),
and references therein.

\bibitem{mos-sidis}
  R.~Meng, F.~I.~Olness and D.~E.~Soper,
  Phys.\ Rev.\  D {\bf 54}, 1919 (1996)
  [arXiv:hep-ph/9511311];
  Nucl.\ Phys.\  B {\bf 371}, 79 (1992).

\bibitem{rr-thesis}
R.~A.~Rodriguez-Pedraza, Ph.D. dissertation,
Iowa State University.

\bibitem{msu-resum}
  F.~Landry, R.~Brock, G.~Ladinsky and C.~P.~Yuan,
  Phys.\ Rev.\  D {\bf 63}, 013004 (2000)
  [arXiv:hep-ph/9905391];
  F.~Landry, R.~Brock, P.~M.~Nadolsky and C.~P.~Yuan,
  Phys.\ Rev.\  D {\bf 67}, 073016 (2003)
  [arXiv:hep-ph/0212159].

\bibitem{qz-resum}
  J.~W.~Qiu and X.~F.~Zhang,
  Phys.\ Rev.\ Lett.\  {\bf 86}, 2724 (2001)
  [arXiv:hep-ph/0012058];
  Phys.\ Rev.\  D {\bf 63}, 114011 (2001)
  [arXiv:hep-ph/0012348].

\bibitem{Berger:1998ev}
  E.~L.~Berger, L.~E.~Gordon and M.~Klasen,
  Phys.\ Rev.\  D {\bf 58}, 074012 (1998)
  [arXiv:hep-ph/9803387].

\bibitem{Qiu:2001ac}
  J.~W.~Qiu, R.~Rodriguez and X.~F.~Zhang,
  Phys.\ Lett.\  B {\bf 506}, 254 (2001)
  [arXiv:hep-ph/0102198].

\bibitem{Berger:2001wr}
  E.~L.~Berger, J.~W.~Qiu and X.~F.~Zhang,
  Phys.\ Rev.\  D {\bf 65}, 034006 (2002)
  [arXiv:hep-ph/0107309].

\bibitem{E772}
  P.~L.~McGaughey {\it et al.}  [E772 Collaboration],
  Phys.\ Rev.\  D {\bf 50}, 3038 (1994)
  [Erratum-ibid.\  D {\bf 60}, 119903 (1999)].

\bibitem{Ji:2004xq}
  X.~D.~Ji, J.~P.~Ma and F.~Yuan,
  Phys.\ Lett.\  B {\bf 597}, 299 (2004)
  [arXiv:hep-ph/0405085];
  Phys.\ Rev.\  D {\bf 71}, 034005 (2005)
  [arXiv:hep-ph/0404183].

\bibitem{tevatron-w-higgs}
  T.~Aaltonen {\it et al.}  [CDF Collaboration],
  arXiv:0707.0085 [hep-ex], Phys. Rev. Lett. (to be published) and 
  references therein.

\bibitem{Nadolsky:1999kb}
  P.~Nadolsky, D.~R.~Stump and C.~P.~Yuan,
  Phys.\ Rev.\  D {\bf 61}, 014003 (1999)
  [Erratum-ibid.\  D {\bf 64}, 059903 (2001)]
  [arXiv:hep-ph/9906280];
  Phys.\ Rev.\  D {\bf 64}, 114011 (2001)
  [arXiv:hep-ph/0012261];
  Phys.\ Lett.\  B {\bf 515}, 175 (2001)
  [arXiv:hep-ph/0012262].

\end{thebibliography}
\end{document}